\documentclass[11pt]{article}
\usepackage{amssymb,epsfig,psfig,graphicx}
\newcommand{\be}{\begin{equation}}
\newcommand{\ee}{\end{equation}}
\newcommand{\bea}{\begin{eqnarray}}
\newcommand{\eea}{\end{eqnarray}}

\begin{document}
\title{\bf Inflation from Tsunami-waves}
\author{{\bf D. Boyanovsky$^{(a,b)}$, F. J. Cao$^{(b)}$, H. J. de
Vega$^{(b,a)}$} \\ \\
(a) Department of Physics and Astronomy, \\ University of
Pittsburgh, \\ Pittsburgh  PA 15260, U.S.A\\(b) LPTHE, Universit\'e
Pierre et Marie Curie (Paris VI) \\ et Denis Diderot
(Paris VII), Tour 16, 1er. \'etage, \\ 4, Place Jussieu, 75252 Paris
cedex 05, France}
\date{\today}
\maketitle
\begin{abstract}
We investigate inflation driven by the evolution of highly
excited {\em quantum states} within the framework of out of
equilibrium field dynamics. These states are characterized by a
non-perturbatively large
number of quanta in a band of momenta but with {\em vanishing}
expectation value of the scalar field. They represent the situation in
which initially a non-perturbatively large energy density is localized
in a band of high energy quantum modes and are  coined tsunami-waves.
The self-consistent evolution of this quantum state and the scale
factor is studied analytically and numerically. It is shown that the
time evolution of these  quantum states lead to two consecutive stages
of inflation under conditions that are the quantum analogue of
slow-roll. The evolution of the scale factor during the first stage
has new features that are characteristic of the quantum state.  During
this initial stage the quantum fluctuations in the highly excited band
build up an effective homogeneous condensate with a non-perturbatively
large amplitude as a consequence of the large number of quanta.
The second stage of inflation is similar to the usual classical
chaotic scenario
but driven by this effective condensate. The excited quantum
modes are already superhorizon in the first stage and do not affect the
power spectrum of scalar perturbations.  Thus, this
tsunami quantum state provides a field theoretical justification
for chaotic scenarios driven by a classical homogeneous
scalar field of large amplitude.
\end{abstract}
\maketitle


\section{Introduction}

A wealth of observational evidence from the temperature anisotropies
in the cosmic microwave background favor inflation as the
mechanism to produce the primordial density
perturbations\cite{turner,cmb}. Thus, inflationary cosmology
emerges as the leading theoretical framework to explain not only the
long-standing shortcomings  of standard big bang
cosmology but also to provide a testable paradigm for structure
formation\cite{revius}-\cite{infl}. The recent explosion in
the quantity and quality of data on temperature anisotropies
elevates inflation to the realm of an experimentally testable scenario
that leads to robust predictions that
withstand detailed scrutiny\cite{turner,cmb}.

However at the level of implementation of an inflationary proposal,
the situation is much less satisfactory. There
are very many different models for inflation motivated by particle
physics and most if not all of them invoke one or several scalar
fields, the inflaton(s), whose dynamical evolution in a scalar
potential leads to an inflationary epoch\cite{revius}-\cite{infl}.
The inflaton field is a scalar field that provides an effective
description for the fields in the grand unified theories. Furthermore
there is the tantalizing prospect of learning some aspects of the
inflationary potential (at least the part of the potential associated with
the last few e-folds) through the temperature anisotropies of the
cosmic microwave background\cite{reconstruction}.

Most treatments of inflation study the evolution of the inflaton via
the {\em classical} equations of motion in the scalar potential
and the effect of quantum fluctuations is typically neglected in the
dynamics of the inflaton. Furthermore, since inflation redshifts
inhomogeneities very fast, the classical evolution is studied in terms
of a {\em homogeneous classical scalar} field. The quantum
field theory interpretation is that this classical, homogeneous field
configuration is the expectation value of a quantum field operator
in a translational invariant quantum state. While the evolution of
this coherent field configuration (the expectation value or
order parameter) is studied via classical equations of motion,
quantum fluctuations of the scalar
field around this expectation value are treated perturbatively and are
interpreted as the  seeds for scalar density perturbations of the
metric\cite{revius}-\cite{infl}.

A fairly broad
catalog of inflationary models based on scalar field dynamics labels
these either as `small field' or
`large field'\cite{reconstruction}. In the `small field' category  the
scalar field begins its evolution with an initial value very near the
origin of the scalar potential and rolls
down towards larger values, an example is new
inflation\cite{revius,coles}. In the `large field' category, the scalar field
begins very high up in the potential hill and rolls down
towards smaller values, an example is chaotic
inflation\cite{revius,coles}.

It is only recently that the influence of quantum fluctuations of the
scalar fields in the dynamical evolution of matter and geometry has
been studied self-consistently, mainly associated with the dynamics of
non-equilibrium phase transitions\cite{noscos,asam} in models
that fall, broadly, in the `small field' category. The conclusion of these
studies is that a  treatment of the quantum fluctuations that couple
self-consistently to the dynamics of the metric
provides a solid quantum field theoretical framework that justifies
microscopically the picture based on classical inflation.
At the same time these studies provide a deeper understanding of the
quantum as well as classical aspects of inflation and inflationary
perturbations. They clearly reveal the classicalization of initial
quantum fluctuations\cite{noscos,asam}, and furnish a microscopic
explanation (and derivation) of the effective, homogeneous classical
inflaton\cite{asam}.

The purpose of this article is to provide a quantum treatment of
models whose classical counterpart are large field models. The {\em classical}
description in these models begins with a homogeneous inflaton scalar with very
large amplitude $\phi \sim M_{Pl}$\cite{revius,coles,infl}, i.e,
very high up
in the scalar potential well. The question of how is this initial condition
achieved is typically answered in terms of a probabilistic
distribution of initial conditions\cite{revius,coles}, as is the case for
chaotic inflation.

Instead,  in this article we study the dynamics that results from the
evolution of a  {\em quantum state} which drives the
dynamics of the scale factor through  the expectation value of the
energy momentum tensor.


The initial state the we consider is a squeezed state with a large
number of particles distributed in a shell in momentum. While
squeezed states had been studied in quantum
optics\cite{quantumoptics} and also in
cosmology\cite{cosmologia} the state considered in this
article is similar to that invoked to model the evolution of
highly excited initial quantum states to describe the dynamics of
heavy ion collisions. These states are characterized by a large
population of quanta localized in momentum shells and had been
coined tsunami-waves in\cite{tsurob,tsu1,tsu2}. The time evolution
of these states leads to the formation of a non-equilibrium plasma
and features properties similar to those expected to occur in the
cooling and expansion of a quark gluon plasma after the
collision\cite{harris}.


{\em The goals of this article:}

We here adapt the ideas and concepts in refs.\cite{tsurob,tsu1,tsu2}
to study the self-consistent
dynamics of the metric and the evolution of a highly excited quantum
state with the goal of providing a quantum description of large field
inflationary models {\em without assuming an expectation
value for the scalar field}. We introduce novel quantum states, which
are the cosmological counterpart of the  tsunami-waves introduced
in\cite{tsurob,tsu1,tsu2} with the following properties:

\begin{itemize}
\item{{\em  Pure states:} the states under consideration, defined by
the wave-functional of the form (\ref{gausswave}) 
are {\em pure states}.  Other proposals involving a
mixed state density matrix as initial state are discussed in
secs. III, IV and the Appendix.  }

\item{{\em  Vanishing expectation value of the scalar field:} Unlike
most models of classical chaotic inflation in which the scalar field 
obtains an expectation value, taken as a classical field, the
expectation value of the scalar field in the 
tsunami-wave states given by the wave-functional (\ref{gausswave})
{\em vanishes}. [See sec. IV  for a non-zero expectation value
of the scalar field]. } 

\item{{\em Highly excited initial modes:} the tsunami-wave
state described by eq. (\ref{gausswave}) with 
the covariance kernel given by eqs. (\ref{covariance}), (\ref{bigomega}) 
describes a  state for which the modes {\em inside} a band are
occupied with a non-perturbatively large ${\cal O}(1/\lambda)$  number
of (adiabatic) quanta. This requires the use of
non-perturbative methods as the large $N$ approach and makes the 
states considered to be very far from the vacuum.
We remark that very high energy modes, 
those that will become superhorizon during the last 10 or so e-folds,
hence are of cosmological importance today, must be in the vacuum
state so as not to lead to  a large amplitude of scalar density
perturbations\cite{revius,infl,noscos}.  } 
\end{itemize}
This type of {\em quantum states} is clearly a novel concept, it
presents an alternative to typical inflationary scenarios that 
invoke the dynamics of a {\em classical} scalar field which in most
cases ignore the quantum dynamics.

The quantum nature of the tsunami states is due to the coherence
between different modes and gives rise
to dynamical consecuences. (See \cite{tsu2}).
In the present case the system is effectively classical  only after the
redshift has assembled the modes into a zero-mode effective
condensate.

The tsunami-wave states described above are the simplest states
and will be the focus of our study. These states can be
generalized to describe mixed-state density matrices and to also
allow for an expectation value of the scalar field. These
generalizations  are described in sections III.C and IV and in the
appendix and are found to lead  qualitatively the same features
revealed by the simpler pure states.

We establish the conditions under which such quantum state leads to
inflationary dynamics and  study in detail the self-consistent
evolution of this quantum state and the space-time metric.

We emphasize that we are {\em not} proposing here yet a new model of
inflation. Instead we focus on  inflation driven by the evolution of a
{\em quantum} state, within the framework of familiar models based
on scalar fields with typical quartic potentials. This is  in contrast
with the usual approach in which the  dynamics
is driven by  the evolution of a homogeneous {\em classical} field of
large amplitude.

{\em Brief summary:} We find that inflation occurs under fairly
general conditions that are the {\em quantum } equivalent of
slow-roll.
There are {\em two} consecutive but distinct inflationary stages: the
first one is completely determined by the quantum features of
the state. Even when the expectation value of the scalar field {\em
vanishes at all times } in this quantum state, the dynamics of the
first stage gives rise to the emergence of an {\em effective classical
homogeneous condensate}. The amplitude of the effective condensate is
non-perturbatively large as a consequence of the non-perturbatively
large number of quanta in the band of excited wavevectors. The second
stage is similar to the familiar classical chaotic scenario, and can be
interpreted as being driven by the dynamics of the effective
homogeneous condensate. The band of excited quantum modes, if not
superhorizon initially they cross the horizon during the first stage
of inflation, hence they do not modify the power spectrum of scalar
density perturbations on wavelengths that are of cosmological
relevance today. Actually, in the explicit examples worked out here,
the excited modes are initially superhorizon due to the generalized
slow-roll condition. Therefore, in a very
well defined manner, tsunami quantum states provide a quantum
field theoretical justification, a microscopic basis, for chaotic
inflation, explaining the classical dynamics of the homogeneous
scalar field.

In section II we introduce the quantum state, obtain the renormalized
equations of motion for the self-consistent evolution of
the quantum state and the scale factor. In section III we  provide
detailed  analytic and numerical studies of the evolution and
highlight the different inflationary stages. In section IV we
discuss generalized scenarios. The summary of results is presented
in the conclusions. We derive the equations of motion for mixed states
in the appendix. 

\section{Initial state and equations of motion}

As emphasized in the introduction, while most works on inflation treat the
dynamics of the inflaton field at the classical level, we use a
quantum description of the inflaton.

We focus on the possibility of inflation through the dynamical {\em
quantum} evolution of  a highly excited initial state with large energy
density. Consistently with inflation at a scale well below the Planck
energy, we treat the inflaton field describing the matter as a
quantum field whereas gravity is treated semiclassically.

The dynamics of the classical space-time metric is determined by the
Einstein equations  with a source term given by the expectation value
of the energy momentum tensor of the quantum inflaton field. The
quantum field evolution is calculated in the resulting metric.

Hence  we solve  {\em self-consistently} the coupled evolution
equations for the classical metric and the quantum inflaton field.

We assume that the universe is homogeneous, isotropic and spatially
flat, thus it is  described by the metric,
\begin{equation}\label{metric}
ds^2 = dt^2 - a^2(t)\; d\vec{x}^2\; .
\end{equation}

Anticipating the need for a non-perturbative treatment of the
evolution of the quantum state, we consider an inflaton
model with an $N$-component scalar inflaton field $\vec{\Phi}(x) $
with quartic self-coupling. We then invoke the
large $N$ limit as a non-perturbative tool to study the
dynamics\cite{noscos,asam,tsu1,tsurob,largeN}.  This choice
is not only motivated by the necessity of a consistent
non-perturbative treatment but also because any
grand unified field theory will contain a large number of scalar
fields, thus justifying a large $N$ limit on more physical grounds.

The matter action and
Lagrangian density are given by
\begin{equation}
S [\vec\Phi] =  \int d^4x\; {\cal L}_m = \int d^4x \;
a^3(t)\left[\frac{1}{2} \, \dot{\vec{\Phi}}^2(x)-\frac{1}{2} \,
\frac{(\vec{\nabla}\vec{\Phi}(x))^2}{a^2(t)}-V(\vec{\Phi}(x))\right]\; ,
\label{action}
\end{equation}
\begin{equation}
V(\vec{\Phi})  =  \frac{m^2}2\; \vec{\Phi}^2 +
\frac{\lambda}{8N}\left(\vec{\Phi}^2\right)^2
+\frac12 \, \xi\; {\cal R} \;\vec{\Phi}^2  \;, \label{potential}
\end{equation}
and we will consider $ m^2 > 0 $, postponing the discussion of the
 case $ m^2 < 0 $ to subsequent work.
 Here $ {\cal R}(t) $ stands for the scalar curvature
\begin{equation}
{\cal R}(t)  =  6\left(\frac{\ddot{a}(t)}{a(t)}+
\frac{\dot{a}^2(t)}{a^2(t)}\right)\; , \label{ricciscalar}
\end{equation}
The $\xi$-coupling of $ {\vec\Phi}^2(x) $ to the scalar curvature $ {\cal
R}(t) $ has been included in the Lagrangian since it is necessary for
the renormalizability of the theory.

The discussion of the alternative inflationary mechanism that we are proposing and the physical description of the
quantum states  becomes more clear in
conformal time
\begin{equation} \label{conftdef}
{\cal T} = \int^t\frac{dt'}{a(t')}
\end{equation}
\noindent in terms of which  the metric is conformal to that in
Minkowski space-time
\begin{equation}
ds^2 = a^2({\cal T})\,(d{\cal T}^2-d{\bf x}^2) \; .
\end{equation}
We introduce the conformally rescaled field
\begin{equation}
\vec\Psi({\cal T}, {\bf x}) = a(t)\,\vec\Phi(t, {\bf x})\label{fieldrescale}
\end{equation}
\noindent in terms of which the  matter action becomes
\begin{equation}
{\cal S}[\vec\Psi] = \int{d{\cal T}\,d^3x\,
\left\{ \frac12\,[(\partial_{{\cal T}}\vec\Psi)^2-
(\nabla\vec\Psi)^2]-a^4({\cal T})\,V\left[\frac{\vec\Psi}{a({\cal T})}\right]+
a^2({\cal T})\,\frac{\cal R}{12}\,\vec\Psi^2 \right\}}\; .
\end{equation}
Since we are interested in describing the time evolution of an initial
quantum state, we pass on to the Hamiltonian description in the
Schr\"odinger representation. This procedure begins by obtaining
the canonical momentum conjugate to the quantum field,
$ \vec\Pi({\cal T}, {\bf x}) $, and the Hamiltonian density
$ {\cal H}({\cal T}, {\bf x}) $
\begin{eqnarray} \label{confhamil}
\vec\Pi({\cal T},{\bf x}) &=& \vec\Psi'({\cal T},{\bf x})\; ,\cr\cr
{\cal H}({\cal T}, {\bf x}) &=& \frac12 \vec\Pi^2
  + \frac12\,(\nabla\vec\Psi)^2
  + a^4({\cal T})\,V\left[\frac{\vec\Psi}{a({\cal T})}\right]
  - a^2({\cal T})\,\frac{\cal R}{12}\,\vec\Psi^2     \; ,     \cr\cr
H({\cal T}) &=& \int{d^3{\bf x} \; {\cal H}({\cal T}, {\bf x}) }\; ,
\end{eqnarray}
where the prime denotes derivative with respect to the conformal time
$ {\cal T} $.

In the Schr\"odinger representation the canonical momentum is given by
\begin{equation}
\Pi^a({\cal T},{\bf x}) = - i \; \frac{\delta\;}
{\delta\Psi^a({\cal T},{\bf x})} \quad ;
\quad a = 1, \ldots, N\; .
\end{equation}
The time evolution of the wave-functional $
\Upsilon\left[\vec{\Psi};{\cal T} \right] $  is obtained from the
functional Schr\"odinger equation
\begin{equation}
i\frac{\partial}{\partial {\cal T}} \Upsilon\left[\vec{\Psi};{\cal T}
\right] = H\left[\frac{\partial}{\partial
\vec{\Psi}};\vec{\Psi}\right] \Upsilon\left[\vec{\Psi};{\cal T}
\right] \label{schroedeqn}
\end{equation}
The implementation of the large $N$ limit begins by writing the field
as follows
\begin{eqnarray} \label{fielddescomp}
\vec\Psi({\bf x},\,{\cal T})
&=& (\sigma({\bf x},\,{\cal T}),\,\vec\pi({\bf x},\,{\cal T}) )\cr\cr
&=& (\sqrt{N}\psi( {\cal T} ) + \chi({\bf x},\,{\cal T}) ,
\,\vec\pi({\bf x},\,{\cal T}) ) \; ,
\end{eqnarray}
\noindent where we choose the `1'-axis in the direction of  the
expectation value of the field and we collectively denote by
$\vec{\pi}$ the $N-1$ perpendicular directions.
\bea
\psi({\cal T}) & = &  \langle\sigma({\bf x},\,{\cal T})\rangle \nonumber \\
\langle\vec\pi({\bf x},\,{\cal T})\rangle & = &  \langle\chi({\bf
x},\,{\cal T})\rangle = 0 \label{fieldsplit} \; ,
\eea
\noindent where the expectations value above are obtained in the state
represented by the wave-functional
 $ \Upsilon\left[\vec{\Psi};{\cal T} \right] $ introduced above.

 The leading order in the large $N$ limit can be efficiently obtained by
 functional methods (see refs.\cite{noscos,asam,tsu1,tsu2,largeN} and
 references therein). The contributions of $ \chi $ to the equations
 of motion are subleading (of order $1/N$) in the large $ N $
 limit\cite{noscos,largeN}.

It is convenient to introduce the spatial Fourier modes of the quantum  field
\begin{equation}
\vec\pi_{{k}}({\cal T}) = \int{d^3x\;\vec\pi({\bf x},\,
{\cal T}) \; e^{i{\bf k}\cdot{\bf x}}}
\end{equation}
\noindent In leading order in the large $N$ limit, the explicit form
of the Hamiltonian is given by\cite{noscos,asam,tsu1,tsu2,largeN}
\begin{eqnarray} \label{hamk}
H({\cal T}) &=& N\, {\cal V}\, h_{cl}({\cal T})
  - \frac{\lambda}{8\,N}\; \left(\sum_{k}
  \langle\vec\pi_{{k}}\cdot\vec\pi_{-{k}}\rangle\right)^2
  + \sum_{k} { H}_{k}({\cal T})\; ,  \cr\cr
h_{cl}({\cal T}) &=&
\frac12\,\psi^{'2}({\cal T})+ \frac{a^2({\cal T})}2\; m^2 \; \psi^2({\cal T}) +
\frac{\lambda}{8}\,\psi^4({\cal T})  ,
\cr\cr
{ H}_{k}({\cal T})  &\equiv&  -\frac{1}{2}\;\frac{\delta^2\;\;}
  {\delta\vec\pi_{{k}}\cdot\delta\vec\pi_{-{k}}}
  + \frac{1}{2} \; \omega^2_k({\cal T})\;
\vec\pi_{{k}}\cdot\vec\pi_{-{k}} \label{hachek} \\ \cr
 \omega^2_k({\cal T}) &\equiv& {k^2} + a^2({\cal T}) \left[{\cal M}^2({\cal
T})-\frac{{\cal R}({\cal T})}{6} \right]
\label{freq2} \; ,\\
{\cal M}^2({\cal T}) &\equiv& m^2 + \xi\,{\cal R}
  + \frac{\lambda}{2}\,\frac{\psi^2}{a^2({\cal T})}
 +\frac{\lambda}{2}\,\frac{\langle\vec\pi^2\rangle}{N\,a^2({\cal
  T})}\label{M2} \; .
\end{eqnarray}
where $ {\cal V} $ is the comoving volume. We assume spherically
symmetric distributions in momentum space.

That is, in the large $N$ limit the Hamiltonian operator (\ref{confhamil})
becomes a time dependent c-number contribution plus a quantum mechanical
contribution, $ \sum_{k} { H}_{k}({\cal T}) $, given by a collection
of harmonic oscillators with time-dependent frequencies, coupled only through
the quantum fluctuations $\langle\vec\pi_{{k}}\cdot\vec\pi_{-{k}}\rangle$.

In eqs.(\ref{hachek})-(\ref{M2}) the scale factor $ a({\cal T}) $ is
determined self-consistently by the Einstein-Friedmann equations.

\subsection{Tsunami initial states}

To highlight the main aspects of the inflationary scenario proposed
here, and to establish a clear difference with
the conventional models, we  now focus our discussion on  the case of
vanishing expectation value, i.e, $\psi({\cal T})=0$,
and a {\em pure quantum state}. The most general
cases with mixed states described by density matrices and
non-vanishing expectation value of the field are
discussed in detail in sections IIIC and  IV and in the appendix.

For $\psi({\cal T})=0$ the quantum Hamiltonian (\ref{hamk}) becomes a
sum of harmonic oscillators with time dependent frequencies that
depend on the quantum fluctuations. Therefore, we propose a Gaussian
wave-functional of the form
\be
\Upsilon\left[\vec{\Psi};{\cal T} \right] = {\cal N}_{\Upsilon}({\cal T})
  \prod_{k}\; e^{-\frac{A_{k}({\cal T})}{2}\;
  \vec\pi_{{k}}\cdot\vec\pi_{-{k}} } \label{gausswave} \; .
\ee
The functional Schr\"odinger equation (\ref{schroedeqn}) in this case
leads to evolution equations for the
normalization factor ${\cal N}_{\Upsilon}({\cal T})$ and the
covariance kernel $ A_k({\cal T})$ whose general
form is found in the appendix (see also \cite{noscos}). The evolution
of the normalization factor is determined by that of $ A_k$, while
the equation for $ A_k$ is
\be
i A'_k({\cal T}) = A^2_k -  \omega^2_k({\cal T}) \label{adoteqn}\; ,
\ee
\noindent where primes refer to derivatives with respect to conformal
time. As described in the appendix for
the general case, the above equation can be linearized
by defining (see appendix)
\begin{equation} \label{defphik2}
 A_{ k}({\cal T}) \equiv - i\,
  \frac{\varphi^{'*}_{k}({\cal T})}{\varphi^*_{k}({\cal T})} \; ,
\end{equation}
where  the mode functions $ \varphi_{k} $ satisfy the equation
\begin{equation} \label{eqmod}
\varphi^{''}_{k} +  \omega^2_ k({\cal T})\; \varphi_{k}= 0
\end{equation}
In terms of these mode functions the self-consistent expectation value
\begin{eqnarray}
\frac{\langle\vec\pi^2\rangle}{N} &=& \frac{1}{N}\int{\frac{d^3k}{(2\pi)^3}\,
  \langle\vec\pi_{{k}}\cdot\vec\pi_{-{k}}\rangle}
\cr\cr \langle\vec\pi_{{k}}\cdot\vec\pi_{-{k}}\rangle
&=& \frac{N}{2A_{R, k}} = {N \over 2 } \; |\varphi_{k}|^2 \; .\label{selfcons}
\end{eqnarray}
We now must provide initial conditions on the wave functional to
completely specify the dynamics. Choosing the initial
(conformal) time at ${\cal T}=0$ with $a({\cal T}=0)=1$, the initial
state is completely specified by furnishing
the real and imaginary parts of the covariance $A_k$ at the initial
time. We parameterize these as\footnote{In the case for which
$ \omega^2_k(0)<0$ we choose $\omega_k(0)=\sqrt{k^2+|{\cal M}^2(0)-{\cal
R}(0)/6|}$.  }
\be
A_{R,k}(0) = \Omega_k ~~; ~~ A_{I,k}(0)= \omega_k(0) \; \delta_k
\label{covariance}
\ee
Choosing the Wronskian of the mode functions $ \varphi_{k}({\cal T}) $ and its
complex conjugate to be
\be
 \varphi_{k}\,\varphi^{'*}_{k} - \varphi^{'}_{k}\,\varphi^*_{k} = 2 i
\label{wronski}
\ee
\noindent determines the following initial condition on the mode
functions (see appendix)
\be
\varphi_{k}(0) = \frac{1}{\sqrt{\Omega_k}} ~~;~~ \varphi'_{k}(0)=
-\left[\omega_k(0)\delta_k +i \; \Omega_k\right]\varphi_{k}(0)\; .
\label{inicondmodes}
\ee
where we also used eqs.(\ref{defphik2}) and (\ref{covariance}).

An important alternative interpretation of these mode functions is
that they form a basis for expanding the
Heisenberg field operators (solution of the Heisenberg equations of motion)
\be
\vec{\pi}(\vec x,{\cal T}) = \int \frac{d^3 k}{(2\pi)^3} \left[
\vec{a}_{k} \; \varphi_{k}({\cal T}) \;e^{i\vec k \cdot \vec x}+
\vec{a}^{\dagger}_{k} \; \varphi^*_{k}({\cal T}) \; e^{-i\vec k \cdot \vec
x}\right] \label{heisenberg} \; ,
\ee
\noindent with $\vec{a}_{k};\vec{a}^{\dagger}_{k}$ annihilation and
creation operators, respectively, with
canonical commutation relations. The Wronskian condition
(\ref{wronski}) ensures that the $ \vec{\pi}(\vec x,{\cal T}) $
fields and their conjugate momenta
obey the canonical commutation relations at equal conformal times.

The physical interpretation of these initial states is highlighted by
focusing on the occupation number
of adiabatic states as well as on the probability distribution of
field configurations.
\begin{itemize}
\item{{\em Occupation number:} it is at this point where the
description in terms of conformal time proves to be
valuable. In conformal time the Hamiltonian in the large $N$ limit is
that of a collection of  harmonic
oscillators with time dependent frequencies. It is then convenient to
introduce the adiabatic occupation number operator
\be
\hat{n}_{k}({\cal T}) = \frac{1}{N}\left[\frac{{ H}_k({\cal T})}{
\omega_k({\cal T})}-\frac{1}{2}\right] \label{ocunumb}  \; ,
\ee
\noindent with $H_k$ given by eq. (\ref{hachek}).

In particular the occupation number at the initial time is given by
(see appendix)
\be
n_k \equiv \langle \hat{n}_{k}(0) \rangle = \frac{\left[\omega_k(0)-\Omega_k
\right]^2+\omega^2_k(0) \;\delta^2_k}{4 \;\omega_k(0) \;\Omega_k}
\label{ocuini} \; .
\ee
Here, the special case with $\Omega_k= \omega_k(0) $ and $ \delta_k=
0$ corresponds to the adiabatic vacuum ($ n_k = 0 $ ). Instead,
we study an initial state in which a band of wave-vectors are
{\em populated with a non-perturbatively large number of
particles}. More precisely,  we consider initial states for which
\bea
n_k &=& {\cal O}\left(\frac{1}{\lambda}\right) \quad
\mbox{inside ~ the ~ excited ~ band}, \cr \cr
n_k &=&   0     \quad   \mbox{outside ~ the ~ excited ~ band.} \; .
\label{band}
\eea
\noindent where $\lambda$ is the quartic self-coupling.

This is accomplished by choosing,
\bea
&& {1 \over \Omega_k} ={\cal O}\left( {1 \over \lambda \; \omega_k(0)}\right)
\quad   \mbox{inside ~ the ~ excited ~ band,} \cr \cr
&&{1 \over \Omega_k} ={1 \over \omega_k(0)}\quad \mbox{and}\quad
\delta_k = 0\quad \mbox{outside ~ the ~ excited ~ band.}\label{bigomega}
\label{lildelta}
\eea
These initial states are highly excited, the expectation
value of the energy momentum tensor in these states leads to an energy
density $\sim 1/\lambda$ and are, therefore, non-perturbative. We
will refer to the case where the excited band is narrow as the
narrow tsunami.

It must be stressed that the particle distribution $ n_k $ alone {\em
partially} determines the initial state. As we see from
eq.(\ref{inicondmodes}), the initial state is completely defined
specifying {\em two} functions of $k : \;  \Omega_k$ and $ \delta_k $ . }

\item{{\em Probability distribution:} an alternative interpretation of
these initial states is obtained by focusing on the
probability distribution of field configurations at the initial
time. It is given by
\be
{\cal P}[\vec{\pi}] = \left|\Upsilon\left[\vec{\Psi};{\cal T}=0
\right]\right|^2 = {\cal N}(0)
  \prod_{k}\; e^{-\Omega_k\;
  \vec\pi_{{k}}\cdot\vec\pi_{-{k}} } \label{probability}
\ee
An intuitive quantum mechanical picture of the wave-functional for the
modes in the excited band is the following. At the
initial time the instantaneous Hamiltonian corresponds to a set of
harmonic oscillators of frequencies $\omega_k(0)$,
while the width in field space of the initial Gaussian state is
determined by $\Omega^{-1/2}_k$.
For a mode in the vacuum state $
1/\sqrt{\Omega_k} \sim 1/\sqrt{\omega_k(0)} $ and the typical
amplitudes of the field are $ {\vec \pi}_{ k} \sim 1/\sqrt{\omega_k(0)}
$ which is the typical width of the potential well. While for a mode
inside the excited band the width in field space
is $ 1/\sqrt{\Omega_k} \sim 1/\sqrt{\lambda \; \omega_k(0)} $ [see
eq.(\ref{bigomega})]. Thus,
large amplitude field configurations with $ {\vec \pi}_{ k\approx k_0}\sim
1/\sqrt{\lambda\; \omega_k(0)} \gg 1/\sqrt{\omega_k(0)} $ have
a probability of ${\cal O}(1)$, i.e, large amplitude configurations
within the band of excited wave-vectors are {\em not} suppressed.
That is, the width of the probability distribution for these modes is
much larger than the typical size of the potential well and there is a
non-negligible probability for finding field configurations with large
amplitudes of ${\cal O}(1/\lambda)$.

These highly excited initial states had been previously proposed as
models to describe the initial stages of ultrarelativistic heavy ion
collisions and had been coined `tsunami
waves'\cite{tsurob,tsu1,tsu2}. They represent spherical shells
(in momentum space) with large occupation numbers of quanta,
describing a state with a large energy density with
particles of a given momentum.  }
\end{itemize}

\subsection{Back to comoving time: renormalized equations of motion}

Having set up the initial value problem in terms of the tsunami-wave
initial wave-functionals, the dynamics is now
completely determined by the set of mode equations eqs.(\ref{eqmod}) with
 eqs.(\ref{freq2})-(\ref{M2}) and the initial
conditions eqs.(\ref{inicondmodes}). However, in order to establish
contact with more familiar results in the literature,
it is convenient to re-write the equations of motion in comoving
time. This is accomplished by the field rescaling
given by eq.(\ref{fieldrescale}) which at the level of mode function
results in introducing the comoving time mode functions
$f_k(t)$ related to the conformal time ones $\varphi_k({\cal T})$ as
\be
f_k(t) = \frac{\varphi_k({\cal T})}{a(t)} \label{comomodes}
\ee
The equations of motion in comoving time for these mode functions are
\begin{eqnarray}
&&\ddot f_{ k}(t) + 3\,H(t)\,\dot f_{ k}(t) + \left[
  \frac{k^2}{a^2(t)} + {\cal M}^2(t)
  \right] f_{ k}(t) = 0 \label{comoeqn}\\
&&{\cal M}^2(t) = m^2 + \xi\,{\cal R}(t)  +
  \frac{\lambda}{4}\,\int\frac{d^3k}{(2\pi)^3}\,  |f_k(t)|^2
  \label{M2como}\\
&& f_{ k}(0) =  \frac{1}{\sqrt{\Omega_{k}}} \;; \quad \quad
\dot f_{ k}(0) =  - [\omega_k(0)\,\delta_{ k} + H(0)
+ i \Omega_{ k}] \; f_{ k}(0) \label{inicondcomo} \; .
\end{eqnarray}
The Einstein-Friedmann equation are,
\be
H^2(t)=\left(\frac{\dot{a}(t)}{a(t)}\right)^2 = \frac{8 \pi
\rho_0}{3 M^2_{Pl}} \quad , \quad \rho_0= \langle T_{00}\rangle
\label{einstein} \; ,
\ee
\noindent where the expectation value is taken in the time evolved
quantum state.
 It is straightforward to see that the expectation value of the
energy-momentum tensor   has the perfect fluid
form, as a consequence of the homogeneity and isotropy of the
system\cite{noscos,asam}.

Thus the set of equations (\ref{comoeqn})-(\ref{einstein}) provide a
closed set of self-consistent equation
for the dynamics of the quantum state and the space-time metric.

\subsubsection{Renormalized Equations of Motion in the Large $ N $ limit}

The  set of equations that determine the dynamics of the quantum state
{\em and} the scale factor
need to be renormalized. The field quantum fluctuations
$$
\frac{\langle \vec \pi^2 \rangle}{N} = \int \frac{d^3k}{2(2\pi)^3}
\; |f_k(t)|^2 \; ,
$$
\noindent requires  subtractions which are absorbed in a
renormalization of the mass, coupling to the Ricci scalar and coupling
constant. The expectation value of the stress tensor also requires
subtractions (but not multiplicative renormalization). Since the
divergence structure is determined by the large energy, short distance
behavior, the
band of excited modes does not influence the renormalization
aspects. Therefore, we use the extensive work on
 the renormalization program which is  available in the literature
referring the reader to  references\cite{noscos,asam}
for details. We here summarize the  aspects that are most relevant for
the present discussion.

First, it is convenient to introduce the following dimensionless quantities,
\begin{eqnarray}
&&\tau = m \; t \quad ; \quad h(\tau)= \frac{H(t)}{m} \quad ;
\quad q=\frac{k}{m} \quad ;
\nonumber \\&&
\omega_q = \frac{\omega_k}{m} \quad ; \quad
\Omega_q = \frac{\Omega_k}{m} \quad ; \quad g= \frac{\lambda}{8\pi^2}
\quad ; \quad
f_q(\tau) = \sqrt{m} \; f_k(t) \; ,
\label{dimvars1}
\eea
\noindent where $ m $ and $ \lambda $ stand for the renormalized mass of the
inflaton and the renormalized self-coupling, respectively\cite{noscos}.
In terms of these dimensionless quantities we now introduce the dimensionless
and fully renormalized expectation value of the self-consistent field as
\bea
&& g\Sigma(\tau) \equiv  \frac{\lambda}{2m^2}\; \langle \pi^2(t)
\rangle_R \nonumber \\
&& \Sigma(\tau)= \int_0^{\infty} q^2 dq \left[ | f_q(\tau)|^2 - {1 \over
{q\; a(\tau)^2}} + {{\Theta(q - 1)}\over {2 q^3}} \left(\frac{{\cal
M}^2(\tau)}{m^2}-{{{\cal{R}(\tau)}}\over{6 m^2}}\right)\right] \;
.\label{sigre}
\eea
\noindent where the terms subtracted inside the integrand renormalize
the mass, coupling to the Ricci scalar and
the coupling constant\cite{noscos,asam}. The dimensionless and renormalized
expressions for the energy density $\epsilon$ and pressure $p$ are given by
\bea
&&\epsilon(\tau)  \equiv {\lambda \over 2 N \; m^4} \langle
T^{00}\rangle_R =
\cr \cr
&&=     \frac{g\Sigma(\tau)}{2} + \frac{[g\Sigma(\tau)]^2}{4} +
\frac{g}{2}\int q^2 \; dq \left\{|\dot{f_q}(\tau)|^2 -
S_1(q,\tau)
+\frac{q^2}{a^2(\tau)} \left[|f_q(\tau)|^2 - S_2(q,\tau)\right]
\right\} \label{energydens} \cr \cr
&& p(\tau)  \equiv {\lambda \over 2 N \; m^4} \, <T^{ii}>_R  \nonumber \\
&& (p+\epsilon)(\tau)   =
g \int q^2 dq \left\{ |\dot{f_q}(\tau)|^2 - S_1(q,\tau)
+\frac{q^2}{3a^2(\tau)}\left[  |f_q(\tau)|^2 - S_2(q,\tau) \right] \right\}
\;. \label{pmase}
\end{eqnarray}
Where the renormalization subtractions $ S_1 $ and  $ S_2 $ are given
by,\cite{noscos,asam}
\begin{eqnarray}
S_1(q,\tau) &=&\frac{q}{a^4(\tau)}+\frac{1}{2qa^4(\tau)}
\left[B(\tau)+2\dot{a}^2  \right] \cr \cr
&+& {\Theta(q - 1) \over {8 q^3 \; a^4(\tau) }}\left[ - B(\tau)^2
- a(\tau)^2 {\ddot B}(\tau) + 3 a(\tau) {\dot a}(\tau) {\dot B}(\tau)
- 4 {\dot a}^2(\tau) B(\tau) \right]\;,\cr\cr
S_2(q,\tau) &=& \frac{1}{qa^2(\tau)}- \frac{1}{2q^3 a^2(\tau)}\;B(\tau)
+ {\Theta(q - 1) \over {8 q^5 \; a^2(\tau) }}\left\{  3 B(\tau)^2
+ a(\tau) \frac{d}{d\tau} \left[ a(\tau) {\dot B}(\tau)\right]\right\}\;,\cr\cr
B(\tau) &\equiv& a^2(\tau)\left[1+g\Sigma(\tau)\right] \; .\label{renosubs}
\end{eqnarray}
We choose  here $\xi =0$ (minimal coupling), the renormalization point
$\kappa = |m|$ and $ a(0)=1 $.

In summary, the set of coupled, self-consistent
equations of motion for the quantum state and the scale factor are
\begin{eqnarray}
&& \left[\frac{d^2}{d \tau^2}+3h(\tau)
\frac{d}{d\tau}+\frac{q^2}{a^2(\tau)}+1+g\Sigma(\tau)
\right]f_q(\tau)  =  0 \label{modknr}  \\
&& f_q(0)  =  \frac{1}{\sqrt{\Omega_q}} \quad ; \quad
\dot{f}_q(0)  = - \left[\omega_q \; \delta_q + h(0)
+ i\,\Omega_q \right]f_q(0) \label{condini} \\
&&\omega_q = \sqrt{q^2+\left|1+g\Sigma(0)-\frac{{\cal
R}(0)}{6m^2}\right|}\label{omegaq} \; ,
\end{eqnarray}
plus  the Einstein-Friedmann equation of motion for the scale factor
\begin{equation} \label{h2tau}
h^2(\tau) = L^2 \, \epsilon(\tau) \qquad ,\qquad
\mbox{where } L^2 \equiv \frac{16 \, \pi N \, m^2}{3\, M^2_{Pl}\, \lambda}
\; . \end{equation}
\noindent with $g\Sigma(\tau)$ and $\epsilon(\tau)$ given by
eqs. (\ref{sigre}) and (\ref{energydens}) respectively.

In order to implement the numerical analysis of the set of
eqs. (\ref{modknr})-(\ref{condini}), (\ref{sigre}) and (\ref{h2tau}) we
introduce an ultraviolet  momentum cutoff $ \Lambda $. For the cases
considered in this article we choose $ \Lambda \sim 200 $ and found
almost no dependence on the cutoff for larger values.
As befits a scalar inflationary model,  the scalar self-coupling is
constrained by the amplitude of
scalar density perturbations  to be $\lambda \sim
10^{-12}$\cite{revius,lyth} implying that   $ g < 10^{-13} $.
Therefore, the  subtractions
can be neglected because $ S_i \sim O(g\Lambda^4) < 10^{-4} $.

The initial state is defined by specifying the $\Omega_q$ and
$\delta_q$. We determine the range of these parameters $\Omega_q$ and
$\delta_q$ by the excitation spectrum for the tsunami-wave initial
state, as well as the condition that lead to inflationary
stage. This will be studied in detail in the next section.

\section{Tsunami inflation}

As emphasized in the previous section, the scenario under
consideration is very different from the popular treatments of
inflation based on the evolution of {\em classical } scalar inflaton
 field\cite{cmb,revius,lyth,infl}. In these scenarios all of the
initial energy is assumed to be in a zero mode (or
order parameter) at the beginning of inflation and the quantum
fluctuations are taken to be perturbatively small with
a negligible contribution to the energy density and the evolution of
the scale factor.

In contrast to this description, our proposal highlights the dynamics
of the {\em quantum states} as the driving mechanism for inflation.
The initial quantum states under consideration correspond to a band
of quantum modes in highly excited states, thus the name
`tsunami-wave'\cite{tsurob,tsu1,tsu2}. This initial state
models a cosmological initial condition in which the energy density
is non-perturbatively large, but concentrated in the
quanta rather than in a zero mode.

We now study under which general conditions such a state can lead to a
period of inflation that satisfies the cosmological constraints for
solving the horizon and entropy problems entailing the necessity for about
60 e-folds of inflation.

It is understood that inflation takes place whenever the expansion of
the universe accelerates, i.e,
\be
 \frac{\ddot a}{a} = h^2 + \dot h = -\frac{L^2}{2}  [\epsilon+3p] > 0\; ,
\label{infla}
\ee
\noindent with $L$ given in eq. (\ref{h2tau}) and $\epsilon$ and $p$
given by eqs. (\ref{energydens}).

While our full analysis rely on the numerical integration of the above
set of equations, much we learn by considering the {\em narrow tsunami
case}.

\subsection{Analytical study: the narrow tsunami case}

Before proceeding to a full numerical study of the equations of
motion, we want to obtain an
analytic estimate of the conditions under which a tsunami initial
quantum state would lead to inflation.

Our main criterion for such initial state to represent high energy
excitations is that the number of quanta in the band of excited
modes is of ${\cal O}(1/\lambda)$.
This criterion, as explained above, is tantamount to requiring that
field configurations with non-perturbative amplitudes
have non-negligible functional probability. Progress can be made
analytically by focusing on the case in which the
band of excited field modes is {\em narrow} i.e, its width $\Delta k $
is such that $\Delta k \ll  k_0$ or in terms of dimensionless quantities
$\Delta q / q_0 \ll 1$. We introduce the following smooth distribution
\be
\Omega_q   =    \frac{\omega_q}{1+ \frac{{\cal N}_{\Omega}}{g}\;
e^{-\left[\frac{q-q_0}{\sqrt{2}\Delta q} \right]^2}} \quad , \quad
\mbox{with}\quad  \frac{\Delta q}{q_0} \ll 1\; ,
\label{distributions}
\ee
\noindent with $\omega_q$ given by eq. (\ref{omegaq}) and ${\cal
N}_{\Omega}$ a normalization constant that fixes the value of the
total energy.

In addition, we choose $  \delta_q  = - h(0)/\omega_q $ as we
discuss below in eq.(\ref{slowrolldelta}).

This initial distribution posses the main features of the
tsunami state described in the previous section.
Since $g \ll 1$, we have for $q \sim q_0$,
\be
\frac{1}{\Omega_q} \sim \frac{1}{g}\gg 1 ~~ \Rightarrow ~~ n_q \sim
\frac{1}{g} \gg 1 \label{biggy}
\ee
\noindent corresponding to highly excited states. While for $|q-q_0|
\gg \Delta q$
\be
\frac{1}{\Omega_q} \sim \frac{1}{\omega_q} ~~ \Rightarrow ~~ n_q \sim 0 \; .
\ee
\noindent Thus, these modes are in a  quantum state near the conformal
(adiabatic) vacuum at the initial time, with
$n_q$ the number of quanta defined by eq. (\ref{ocuini}) in terms of
dimensionless variables. For these
distributions (narrow tsunamis),  the integral over mode functions for
the quantum fluctuations $g\Sigma(\tau)$  [given by eq. (\ref{sigre})]
is dominated by the narrow band of excited states with mode amplitudes
$\sim 1/\sqrt{g}$ and can be approximated by
\begin{equation} \label{apgsigI}
g\Sigma(\tau) = g\; \Delta q \; q_0^2\;  |f_{q_0}(\tau)|^2 + O(g)
+ O(g\,\Delta q) \simeq |\phi_{q_0}(\tau)|^2 \; ,
\end{equation}
where we have introduced the effective $q_0$ mode
\be
 \phi_{q_0}(\tau) \equiv \sqrt{g \; \Delta q} \; q_0 \;
f_{q_0}(\tau). \label{effzeromode}
\ee
\noindent we note that the initial condition (\ref{condini})  and the
tsunami-wave condition (\ref{biggy}) entail that
despite the presence of the coupling constant in its definition, the
amplitude of the effective $q_0$ mode is of ${\cal O}(1)$.

The equation of motion for the effective $q_0$-mode takes the form
\begin{equation} \label{modoq0}
{\ddot \phi}_{q_0}(\tau) + 3 \, h(\tau) \, {\dot \phi}_{q_0}(\tau) + \left[
{q_0^2 \over a^2(\tau)} + 1 + |\phi_{q_0}(\tau)|^2
\right]\phi_{q_0}(\tau) = 0 \; .
\end{equation}
The scale factor follows from
\begin{equation} \label{friedq0}
h^2(\tau) = L^2 \; \epsilon(\tau) \; ,
\end{equation}
with energy and pressure,
\begin{eqnarray} \label{eypq0}
\epsilon(\tau) &=& \frac12 \, |{\dot \phi}_{q_0}(\tau)|^2
+ \frac12 \, |\phi_{q_0}(\tau)|^2 + \frac14 \; |\phi_{q_0}(\tau)|^4
+ \frac{q_0^2}{2 \, a^2(\tau)} \; |\phi_{q_0}(\tau)|^2 \; ,\cr \cr
(p+ \epsilon)(\tau) &=& |{\dot \phi}_{q_0}(\tau)|^2 +
\frac{q_0^2}{3 \, a^2(\tau)} \; |\phi_{q_0}(\tau)|^2 \; .
\end{eqnarray}
\noindent where we have neglected terms of ${\cal O}(g)$.

We will refer to the set of evolution equations
(\ref{modoq0})-(\ref{eypq0}) as the one mode approximation evolution equations.

In particular, within this one-mode approximation, the acceleration of
the scale factor obeys
\begin{equation}
\frac{\ddot a(\tau)}{a(\tau)} = - L^2 \left[|{\dot \phi}_{q_0}(\tau)|^2 - \frac{|\phi_{q_0}(\tau)|^2}{2}-  \frac{|\phi_{q_0}(\tau)|^4}{4}  \right] \label{acc}
\end{equation}

Therefore, the condition for an inflationary epoch,  $ \ddot a > 0 $, becomes
\begin{equation}
|{\dot \phi}_{q_0}(\tau)|^2 \quad
< \quad \frac12 \, |\phi_{q_0}(\tau)|^2 + \frac14 \; |\phi_{q_0}(\tau)|^4 \;.
\end{equation}
A sufficient criterion that guarantees inflation is the {\em
tsunami slow roll condition}
\be
 |{\dot \phi}_{q_0}(\tau)|  \ll
|\phi_{q_0}(\tau)| \label{slowroll}
\ee
The initial conditions (\ref{condini}) and the condition that
$\Omega_{q_0} \sim g \ll 1$ imply that the tsunami
slow roll condition (\ref{slowroll}) at early times is guaranteed if
$\delta_{q_0}$ is such that
\be
|\omega_{q_0}\delta_{q_0} + h(0)| \ll 1 \label{slowrolldelta}
\ee
Hence tsunami-wave initial states that satisfy the tsunami
slow-roll condition (\ref{slowrolldelta}) {\em lead
to an inflationary stage}.

Moreover,  in order to have slow roll (\ref{slowroll}) at later times,
the effective friction coefficient $ 3 \, h(\tau) $  should be larger
than the square of the frequency in the evolution equations
(\ref{modknr}). That is,
\begin{equation}\label{rueda}
\frac{q_0^2+1+g\Sigma(0)}{3 h(0)} \ll 1 \;.
\end{equation}
(\emph{i.e.} the $q_0$-mode should be deep inside the overdamped
oscillatory regime). Eq.(\ref{rueda})  implies that $ h(0) \gg 1 $ and
this together with eq.(\ref{slowrolldelta}) implies  that $ \delta_{q_0}
$ must be negative.

A remarkable aspect of the narrow tsunami state is that it leads to a
dynamical evolution
of the metric similar to that obtained in {\em classical chaotic inflationary
scenarios} in the slow roll approximation\cite{revius,lyth}. In
particular the expression for the acceleration (\ref{acc}) and the
tsunami slow roll condition (\ref{slowroll}) are indeed similar to
those obtained in  classical chaotic inflationary models driven by a
homogeneous
classical field (zero mode). However, despite the striking similarity with
 classical chaotic models, we haste to add that both the
conditions that define a tsunami state and the tsunami slow roll
condition (\ref{slowroll}) guaranteed by the initial
value (\ref{slowrolldelta}) is of purely quantum mechanical origin in
contrast with the classical chaotic-slow-roll
scenario. Furthermore, we recall that the expectation value of the
scalar field vanishes in this state.

\subsubsection{Early time dynamics} \label{tsuearly}

Under the assumption of a tsunami wave initial state
and the tsunami slow-roll condition (\ref{slowroll})
the contribution  $ {\dot \phi}_{q_0}(\tau) $  in the energy and
in the pressure [see eq. (\ref{eypq0})] can be neglected  provided,
\begin{equation} \label{estit}
|{\dot \phi}_{q_0}(\tau)|^2 \ll \frac{q_0^2}{3 \, a^2(\tau)} \;
 |\phi_{q_0}(\tau)|^2 \; .
\end{equation}
We call $ \tau_A $ the time scale at which this rely no longer holds.
Furthermore, we can approximate $ \phi_{q_0}(\tau) $ by $
\phi_{q_0}(0) $ if
\begin{equation} \label{estit2}
\tau_A \ll \left| { \phi_{q_0}(0) \over {\dot \phi}_{q_0}(0)} \right| \; .
\end{equation}
This condition is fulfilled at least for $ \tau_A \lesssim 1 $ due to
the tsunami slow-roll condition (\ref{slowroll}).

During this interval the Friedmann equation (\ref{friedq0}) takes  the form,
\begin{equation} \label{friango}
\left[ {\dot a}(\tau) \over a(\tau) \right]^2
=  L^2 \; \left[ \frac12 \, |\phi_{q_0}(0)|^2 + \frac14 \; |\phi_{q_0}(0)|^4
+ \frac{q_0^2}{2 \, a^2(\tau)} \; |\phi_{q_0}(0)|^2 \right]
= \frac{D}{a^2(\tau)} + E \; .
\end{equation}
where we used that $ g \Sigma(0) = |\phi_{q_0}(0)|^2 $.
This equation is valid as long as the characteristic time scale of variation of the
metric  is  shorter than that  of the mode
$ \phi_{q_0}(\tau) $.

The preceding equation can be integrated with solution
\begin{eqnarray} \label{soltaua}
a(\tau) &=& \sqrt{D \over E} \, \sinh\left(\sqrt{E}\, \tau
+c\right) \quad , \quad \frac{\ddot a(\tau)}{a(\tau)} =  E \;>\; 0 \;,
\cr\cr
h(\tau) &=& \sqrt{E} \, \coth\left(\sqrt{E}\, \tau  + c\right)
 \quad , \quad
{\dot h}(\tau) = - { E \over \sinh^2\left(\sqrt{E}\, \tau
+c\right)} \; , \label{tsuni}
\end{eqnarray}
where the constants $ D $, $ E $ and $ c $ are given by,
\begin{equation} \label{ctestaua}
D = L^2 \; \frac{q_0^2}{2} \; g\Sigma_0 \quad , \quad
E = L^2\;\left( \frac{g\Sigma_0}{2} + \frac{g\Sigma_0^2}{4}\right) \quad,\quad
\sinh c = \sqrt{\frac{E}{D}} \; ,
\end{equation}
and $ g\Sigma_0 \equiv g \Sigma(0) = |\phi_{q_0}(0)|^2 $.

We see from eqs. (\ref{soltaua}) that during this interval there is  an
inflationary stage with an accelerated expansion $ \frac{\ddot
a(\tau)}{a(\tau)} = E > 0 $.
We also see that $ h(\tau) $ decreases with time until it reaches the constant
value $ \sqrt{E} $ that determines the onset of a quasi-De Sitter inflationary
stage.

We  now estimate the range of validity of the solution in
eqs. (\ref{soltaua}). The first condition in eq. (\ref{estit}) is more
stringent than the second one in eq. (\ref{estit2}) for most of the
interesting range of parameters.

 $ \tau_A $ determines  the time scale  at which the solution (\ref{soltaua})
ceases to be valid, and the condition (\ref{estit}) is no longer
fulfilled, i.e,
\begin{equation} \label{condtaua}
|\dot\phi_{q_0}(\tau_A)|^2 \sim \frac{q_0^2}{3\,a^2(\tau_A)} \;
|\phi_{q_0}(\tau_A)|^2 \; .
\end{equation}
When this equation is valid, we see from eq. (\ref{eypq0}) that $ p +
\epsilon  $ becomes of the order of $ |\dot\phi_{q_0}(\tau_A)|^2
$. Furthermore, the above condition together with the slow roll condition,
eq. (\ref{slowroll}), leads to
\begin{equation} \label{aplas}
\frac{q_0^2}{a^2(\tau_A)} \ll 1 \; .
\end{equation}
Therefore, from eq.(\ref{friango}) we see that $h(\tau_A) \sim
\sqrt{E}$ and is slowly varying.

Since the slow roll condition guarantees that $
|\ddot\phi_q(\tau_A)| \ll |\phi_q(\tau_A)| $, we can now use
eq.(\ref{apgsigI}) along with the evolution equation
(\ref{modoq0}) which setting $ h =
\sqrt{E} $ leads to the following relation
\begin{equation} \label{fiptoA}
\dot\phi_{q_0}(\tau_A) \simeq - \frac{1+g\Sigma_0}{3\sqrt{E}}\;
\phi_{q_0}(\tau_A) \; .
\end{equation}
From eq.(\ref{condtaua}) the scale $\tau_A$
is therefore estimated to be given by
\begin{equation} \label{taua}
\tau_A \sim {1 \over \sqrt{E}}\left[ \mbox{ArgSinh}
\left(\frac{\sqrt{3} \, q_0 \, E}
{\sqrt{D}(1+g\Sigma_0)} \right)\;-\;c \right]
= {1 \over \sqrt{E}}\left[ \mbox{ArgSinh} \left(
\frac{L\sqrt{3g\Sigma_0}\, (1+g\Sigma_0/2)}{\sqrt{2}(1+g\Sigma_0)}
\right)\;-\;c
\right]\;.
\end{equation}
This initial inflationary period with a  decreasing Hubble
parameter exists provided the r.h.s. is here positive, {\it i.e.},
\begin{equation} \label{q0condtaua}
q_0 > \frac{1+g\Sigma_0}{\sqrt{3\,E}} \; .
\end{equation}
In order to distinguish this phase from the later stages, to be
described below, we
refer to this early time inflationary stage as `tsunami-wave
inflation' because the
distinct evolution of the scale factor during this stage is
consequence of the tsunami-wave properties.

At $ \tau = \tau_A $ we have:
\begin{eqnarray} \label{valuesattaua}
\phi_{q_0}(\tau_A) \simeq \phi_{q_0}(0) = \sqrt{g\Sigma_0} \qquad &,&  \qquad
\dot\phi_{q_0}(\tau_A) \simeq  - \frac{1+g\Sigma_0}{3\,h(\tau_A)}
\; \phi_{q_0}(\tau_A) \;, \cr\cr
a(\tau_A) \simeq \frac{\sqrt{3\,E}\,q_0}{1+g\Sigma_0} \qquad &,&
\qquad h(\tau_A) \simeq  \sqrt{E} \; .
\end{eqnarray}

For $ \tau  >  \tau_A $, $ q_0/a(\tau) <<1 $ and the physical
wavevectors in the excited band have red-shifted so
much that all  terms containing $ q_0 $ become negligible in the
evolution equations. Therefore, all modes in the excited band evolve
as an {\em effective}  $ q= 0 $ mode.
Hence for $\tau > \tau_A$ the dynamics of the scale factor is
described by an effective homogeneous zero mode and
describes a different regime from the one studied above. Such regime is akin to
the classical chaotic scenario.

\subsubsection{The  effective classical chaotic inflationary
epoch}\label{tsulate}

For $\tau >>\tau_A$, when $ q_0^2/a^2 \ll |\dot\phi_q|^2/|\phi_q|^2 \ll 1 $
 all the physical momenta corresponding to the comoving wavevectors in
 the excited band
have redshifted to become negligible in the equations of motion. The
 dynamics is now  determined by the following
set of equations for the effective zero mode and the scale factor,

\begin{eqnarray} \label{evoleqtaua}
&&{\ddot \phi}_{q_0}(\tau) + 3 \, h(\tau) \, {\dot \phi}_{q_0}(\tau) +
\left[1 + |\phi_{q_0}(\tau)|^2\right]\phi_{q_0}(\tau) = 0 \; , \cr \cr
&&h^2(\tau) = L^2 \; \epsilon(\tau) \; ,
\end{eqnarray}
where the energy and  pressure are given by,
\begin{eqnarray} \label{petaua}
\epsilon(\tau) &=& \frac12 \, |{\dot \phi}_{q_0}(\tau)|^2
+ \frac12 \, |\phi_{q_0}(\tau)|^2 + \frac14 \; |\phi_{q_0}(\tau)|^4 \; ,\cr \cr
(p+ \epsilon)(\tau) &=& |{\dot \phi}_{q_0}(\tau)|^2 \; .
\end{eqnarray}

The initial conditions on $\phi_{q_0}$ and ${\dot\phi}_{q_0}$ are
determined by their values at the time $\tau_A$, while the
slow-roll condition (\ref{slowroll}) determines that the {\em
imaginary} parts of $ \phi_{q_0} $ and $ \dot\phi_{q_0} $ are
negligible.

Therefore, after $ \tau_A $ the dynamic is identical to that of a classical homogeneous
field (zero mode)
\begin{equation}
\eta_{eff}(\tau) = \mbox{Re}[\phi_{q_0}(\tau)] \; ,
\end{equation}
that satisfies the equations of motion,
\begin{eqnarray}
&&\ddot\eta_{eff} + 3\,h\,\dot\eta_{eff} +(1+\eta_{eff}^2)\,\eta_{eff} = 0\;,
\cr\cr
&&h^2(\tau) = L^2 \; \epsilon(\tau) \; . \label{classy}
\end{eqnarray}
with energy and pressure,
\begin{eqnarray} \label{peeff}
\epsilon(\tau) &=& \frac12 \, \dot\eta_{eff}^2 + \frac12 \, \eta_{eff}^2
+\frac14 \; \eta_{eff}^4 \; ,\cr \cr
(p+ \epsilon)(\tau) &=& \dot\eta_{eff}^2 \; , \label{enermo}
\end{eqnarray}
and initial conditions [using eq. (\ref{valuesattaua})],
\begin{eqnarray} \label{magtaua}
\eta_{eff}(\tau_A) &=& \phi_{q_0}(\tau_A) = \phi_{q_0}(0) = \sqrt{g\Sigma_0}
\; , \cr \cr
\dot\eta_{eff}(\tau_A) &=& \dot\phi_{q_0}(\tau_A) =
- \frac{1+g\Sigma_0}{3\,h(\tau_A)} \; \eta_{eff}(\tau_A) \; .
\end{eqnarray}
Where the value of $ \dot\eta_{eff}(\tau_A) $ is determined by
the slow roll condition ($ \Rightarrow  \ddot\phi_{q_0}(\tau_A) \simeq 0 $),
the evolution eq. (\ref{modoq0}) and eq. (\ref{aplas}) and $a(\tau_A)$
and $h(\tau_A)$ are given by eq. (\ref{valuesattaua}).

When $ g\Sigma_0 \ll 1 $, the quadratic term in the potential dominates, and
we can integrate the previous equations to obtain
\begin{equation}
\eta_{eff}(\tau) = \eta_{eff}(\tau_A) - \frac{\sqrt{2}}{3L}(\tau-\tau_A) \;,
\quad \mbox{(for $ g\Sigma_0 \ll 1 $)} \;.
\end{equation}

This evolution is similar to that of classical chaotic inflationary
models\cite{revius,lyth}. Therefore
for $\tau > \tau_A$ when the physical momenta in the excited band have
redshifted so much that their contribution
in the equations of motion of the quantum modes and the energy and
pressure become negligible, the evolution of
the quantum modes and the metric is akin to a classical chaotic
inflationary scenario driven by a homogeneous
c-number scalar field. This equivalence allows us to use  the results
obtained for classical chaotic inflation.
Thus, as the classical slow roll condition
($ |\dot\eta_{eff}| \ll |\eta_{eff}| $) holds, the evolution of the
effective scalar field is
overdamped and
the system enters a quasi-De Sitter inflationary epoch.
This inflationary period ends when the slowly decreasing Hubble parameter
becomes of the order of the inflaton mass, i.e,  $ 3\,h \sim 1 +
\eta_{eff}^2 $.
At this stage the effective classical field exits the overdamped
regime and starts
to oscillate, the slow roll condition no longer holds  and a matter dominated
epoch ($ |\dot\eta_{eff}| \sim |\eta_{eff}| \Longrightarrow p \sim 0
$) follows.

However, we emphasize that while the effective zero mode $\eta_{eff}$
obeys a classical
equation of motion and that the components of the energy momentum
tensor eq.(\ref{enermo}) are those
from a classical field, the origin of this mode is {\em purely quantum
mechanical}. From
the identification (\ref{magtaua}) and eq. (\ref{apgsigI}) it is
clear that the effective zero mode
is a collective superposition of modes in the highly excited
band. From the initial and tsunami wave conditions
$f_{q_0}(\tau_A) \sim 1/\sqrt{\Omega_{q_0}} \sim 1/\sqrt{g}$ it
follows that the amplitude of the effective zero mode is
$\eta_{eff}(\tau_A) \sim q_0\sqrt{\Delta q}$. Restoring the dimensions
and the proper powers of the coupling that
were absorbed in the constant $ L $ in the Friedmann equation we find that
the equations of motion (\ref{classy}) with
the stress tensor components (\ref{peeff}) are those obtained from a
(dimensionfull) classical homogeneous
field $\varphi_{eff}(t) $ with a classical potential
$V(\varphi_{eff})$ given by
\bea
&&\varphi_{eff}(t) = \frac{m\, \sqrt2}{\sqrt{\lambda}} \;
\eta_{eff}(\tau) \nonumber \\
&& V(\varphi_{eff}) = \frac{m^2}{2}\;\varphi^2_{eff} +
\frac{\lambda}{8}\;\varphi^4_{eff} \label{effective}
\eea
 \noindent with the initial value at the time $t_A=\tau_A/m$
\be
\varphi_{eff}(t_A) \sim \frac{m}{\sqrt{\lambda}} \sqrt{k^2_0  \; \Delta
k \over m^3} \label{classampli}
\ee
The non-perturbative amplitude of the effective zero mode is a
consequence of the non-perturbative amplitude
of the  excited {\em quantum} modes with an ${\cal O}(1/\lambda)$
number of quanta.
\subsubsection{Number of e-folds}
An important cosmological quantity is the total number of e-folds
during inflation. As discussed above, there are two different
inflationary stages, the first one is determined by the
equations (\ref{friango})-(\ref{tsuni}) and characterized by a rapid
fall-off of the Hubble parameter approaching a quasi-De Sitter
stage. This new stage has been referred to as the
tsunami-wave inflationary stage above to emphasize that the
dynamics is determined by the distinct characteristics of the
tsunami-wave initial stage.

The second stage is described by an effective zero mode and
the evolution equations (\ref{classy}, \ref{peeff})  and is akin to
the chaotic inflationary stage driven
by a classical homogeneous scalar field. The crossover between the two
regimes is determined by the time
scale $\tau_A$ (in units of the inflaton mass) and given by
eqs.(\ref{taua}) at which the contribution from
the term $q^2_0/a^2(\tau)$ to the equations of motion becomes
negligible. Therefore there are to distinct contributions
to the total number of e-folds, which is given by
\begin{equation} \label{nenarrow}
N_e(q_0, h(0)) = \log a(\tau_A) + N_e(0, h(\tau_A)) \; ,
\end{equation}
where $ a(\tau_A) $ is given by eq. (\ref{magtaua}) and
$ N_e(0, h(\tau_A)) $ is just the number of e-folds for classical chaotic
inflation with an initial Hubble parameter $ h(\tau_A) $.

We can express $ h(\tau_A) $ as a function of $ q_0 $ and $ h(0) $,
\begin{equation}
h(\tau_A) = L \; \sqrt{\frac{g\Sigma_0}{2}+\frac{(g\Sigma_0)^2}{4}}
= \sqrt{h^2(0)-L^2\;\frac{q_0^2}{2}\;g\Sigma_0} \;.
\end{equation}
The number of e-folds during the first stage, is given by
\be
\log a(\tau_A) \sim \log\left[\frac{\sqrt{3\,E}\,q_0}{1+g\Sigma_0}
\right] \label{efoldtsunami}
\ee
The expression for the number of e-folds during the following, chaotic
inflationary stage simplifies when $ g\Sigma_0 \ll 1 $.
In this case the quadratic term in the potential dominates, and we can
obtain simple analytical expressions
\begin{equation} \label{lnane0narrow}
N_e(0, h(\tau_A)) = \frac{3L^2}{4} \; \eta_{eff}^2(\tau_A)
= \frac{3L^2}{4} \; g\Sigma_0
= \frac{3L^2}{2} \; \frac{\epsilon_0}{1+q_0^2}     \;,
\quad \quad \quad
\mbox{(for $ g\Sigma_0 \ll 1 $)} \;.
\end{equation}
We see that the number of efolds grow when $ q_0 $ decreases at fixed
initial energy $ \epsilon_0 $. That is, we have more efolds when the energy is
concentrated at low momenta.

\subsubsection{In summary}

Before proceeding to a full numerical study of the evolution we
summarize the main features of the dynamics gleaned from the narrow
tsunami case to compare with the numerical results.

\begin{itemize}

\item{The conditions for tsunami-wave inflation are {\em i) } a band
of excited states centered at a momentum $k_0$
with a non-perturbatively large ${\cal
O}(1/g)$ number of quanta in this band, and {\em ii)} the tsunami
slow-roll condition eq.(\ref{slowroll}). These conditions are
guaranteed by the initial conditions
on the mode functions given by eq.(\ref{condini}) with the tsunami-wave
distributions of the general form given by
eqs. (\ref{distributions}), (\ref{slowrolldelta}). }

\item{
There are two successive inflationary periods. During the  first
one, described in sec. \ref{tsuearly}, the dynamics is completely
characterized by the distinct features of the
tsunami-wave initial state,  the Hubble parameter falls off fast and
reaches an approximately  constant value $ \sqrt{E} $ that
characterizes the quasi-De Sitter epoch of
inflation of the second period. The second stage, described in
sec. \ref{tsulate}  can be described in terms of an effective
classical zero mode and the
evolution of this effective mode and that of the Hubble parameter are
akin to the standard chaotic inflationary scenario.}

\item{
The tsunami-wave initial state can be interpreted as a {\em
microscopic} justification of the classical chaotic scenario described by an
effective classical zero mode of large amplitude. The amplitude of
this effective zero mode is {\em non-perturbative} as
a consequence of the non-perturbative ${\cal O}(1/\lambda)$ number of
quanta in the narrow band of
excited modes. Thus the initial value of the effective, classical zero
mode that describes the second, chaotic inflationary stage, is
completely determined by the quantum initial state.
}

\item{An important point from the perspective of structure formation
is that the band of excited wavevectors centered at $q_0$ either
correspond to superhorizon modes initially, or all of the excited
modes cross the horizon during the first stage of inflation, i.e,
during the tsunami stage. This is important because the chaotic second
stage of inflation which dominates during a longer period guarantees
that the band of excited modes have become superhorizon well before
the last 10 e-folds of inflation and hence cannot affect the
power spectrum of the temperature anisotropies in the CMB. The fact
that the tsunami-wave initial state is such that the very high
energy modes (necessarily trans-Planckian) that cross the horizon
during the last 10 e-folds and are therefore of cosmological
importance today are in their (conformal) vacuum state leads to the
usual results from chaotic inflation for the power spectrum
of scalar density perturbations.   }

\end{itemize}

Although these conclusions are based on the narrow tsunami
case, we will see below that a full numerical integration
of the self-consistent set of equations of motion confirms this picture.

In sections \ref{sotherdist} and \ref{sgenchaoinf} we show how
this results can be easily extended to {\em more general particle
distributions} and {\em more general initial states}.

\subsection{Numerical example}

To make contact with familiar models of inflation with an inflaton
 field with a mass  near the grand unification scale, we choose the
 following  values of the parameters:
\be
 \frac{m}{ M_{Pl}} = 10^{-4}~ ,~
\; \lambda = 10^{-12} ~ , ~ \; N = 20
\ee
\noindent where the number of scalar fields $N=20$ has been chosen as
a generic representative of a grand unified quantum field theory.
 For these values we find
$$
L^2 \equiv \frac{16 \, \pi N \, m^2}{3\, M^2_{Pl}\, \lambda}
= 3.35 \cdot 10^6 \; .
$$

As an example we shall consider an initial energy density
$ \rho_0 = \langle T_{00} \rangle  = 10^{-2} \, M_{Pl}^4 $.
Thus, the initial value for the Hubble parameter is
$ H_0 = \sqrt{8 \pi  \rho_0 / 3M_{Pl}} = 3.53 \cdot 10^{18} \; GeV
(= 1.654 \cdot 10^{52} \; km/s/Mpc) $.
These initial conditions in dimensionless  variables give
$ \epsilon_0 = 2.50 $ and $ h(0) = 2890 $.

In addition, the slow roll conditions (\ref{rueda}) imply:
$$
\frac{q_0^2+1+g\Sigma_0}{3\,h(0)} \ll 1
$$
which  in this case results in
$$
q_0 \ll 95 \;.
$$
We choose $ q_0 = 80.0 $, and initial conditions in
 eq. (\ref{condini}) with  $\Omega_q$ and $\delta_q$ given by
 eq. (\ref{distributions}) and (\ref{slowrolldelta}). These initial conditions
 satisfy the tsunami slow roll condition,
\begin{eqnarray}
|\omega_q\;\delta_q+h(0)| \ll 1
\end{eqnarray}
Furthermore,  we take $ \Delta q = 0.1 $ and $ {\cal N}_\Omega $ is
adjusted by fixing the value $ g\Sigma(0) = g\Sigma_0 $ which for the
values chosen  for $ \epsilon_0 $ and by
eqs.(\ref{energydens}) and (\ref{apgsigI}) and (\ref{slowroll}) gives
$ g\Sigma_0 = 7.81 \cdot 10^{-4} $.

\begin{figure}[ht!]
\epsfig{file=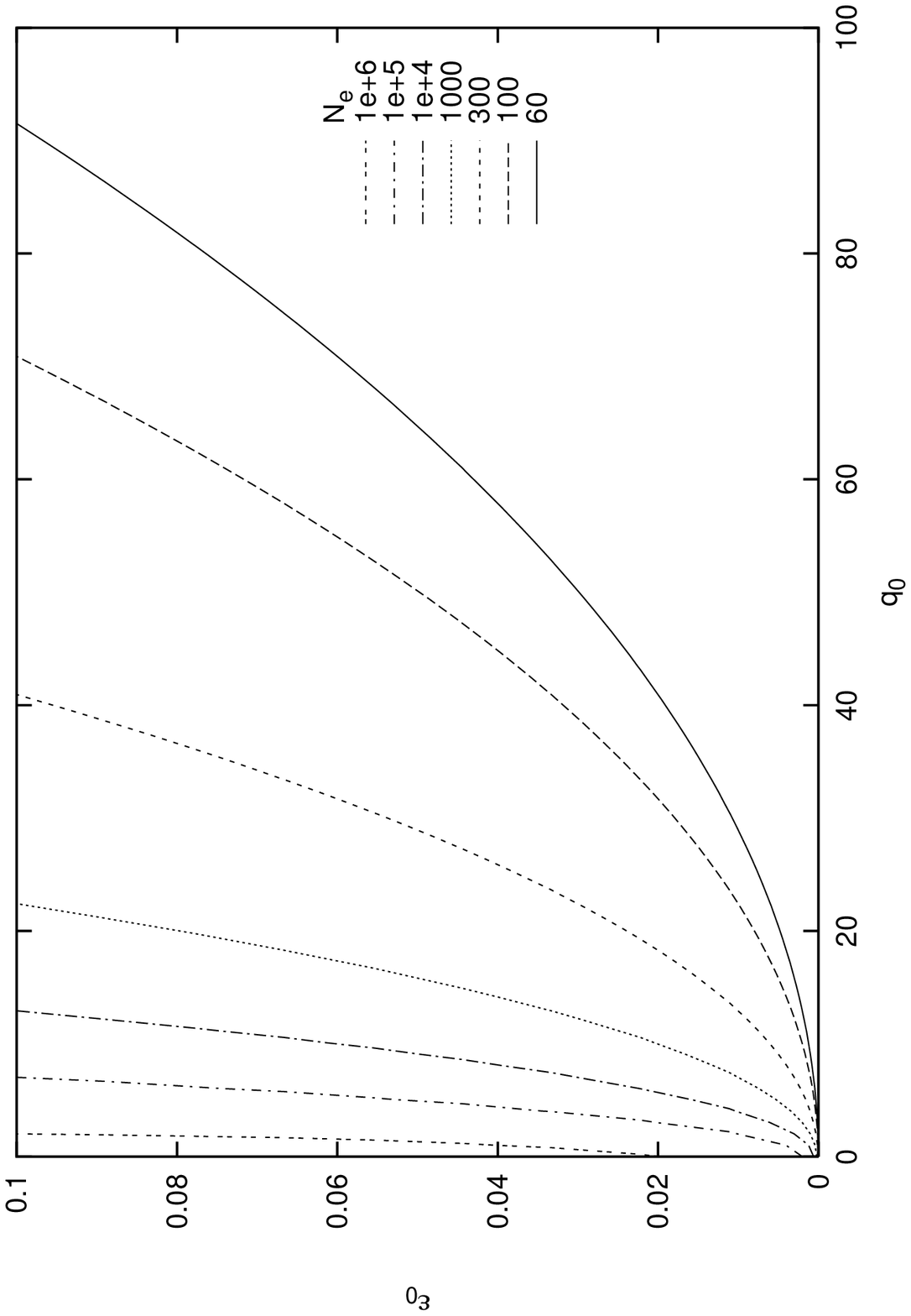,width=6in,height=6in}
\vspace{.1in}
\caption{Tsunami  inflation: isolines of
constant number of efolds obtained from eq.(\ref{lnane0narrow})
(valid for $ g\Sigma_0 \ll 1 $),
for $ m = 10^{-4} M_{Pl} $, $ \lambda = 10^{-12} $ and $ N = 20 $.}
\label{Nefig}
\end{figure}

\begin{figure}[h]
\epsfig{file=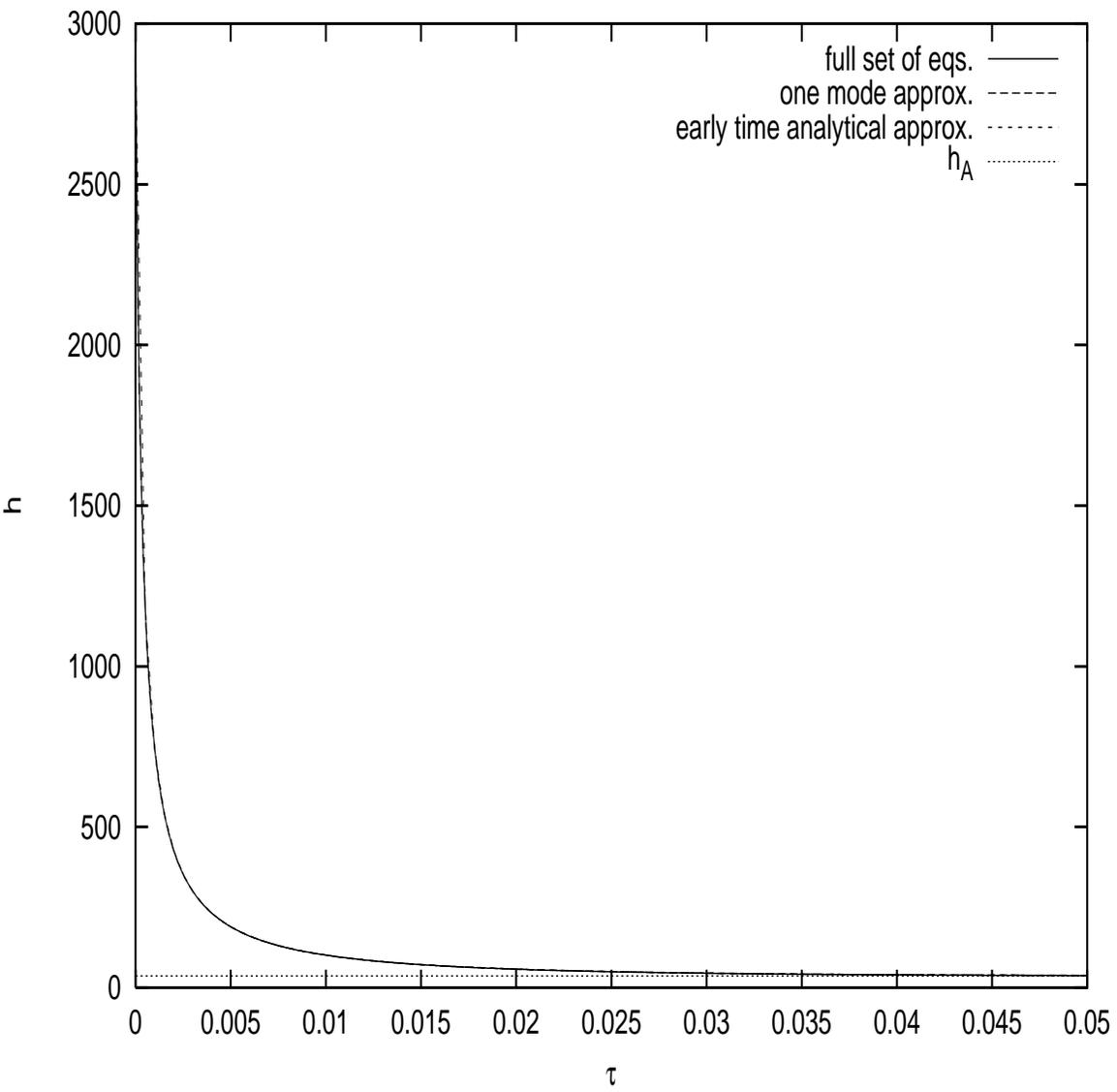,width=6in,height=6in}
\vspace{.1in}
\caption{Tsunami  inflation:
Early time $ h(\tau) $. $ h_A \equiv h(\tau_A)$ is the asymptotic value for
the early period that ends at $ \tau_A \sim 0.133 $.
For $ m = 10^{-4} M_{Planck} $, $ \lambda = 10^{-12} $ and $ N = 20 $.
Initial conditions:
$ \rho_0 = 10^{-2} M_{Pl}^4 $, $ q_0 = 80.0 $ and $ \Delta q = 0.1 q_0 $.}
\label{hearlyfig}
\end{figure}

\begin{figure}[h]
\epsfig{file=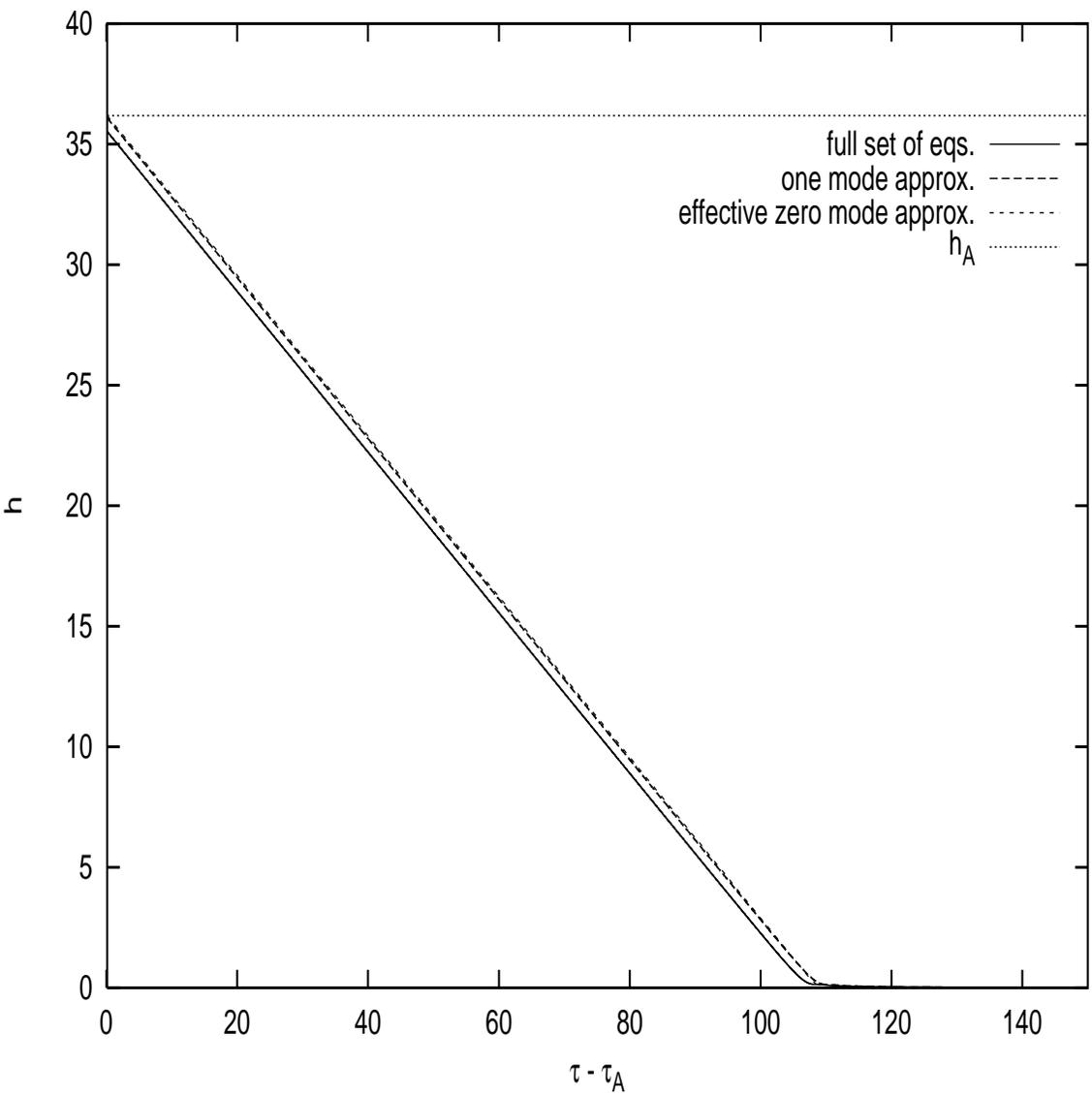,width=6in,height=6in}
\vspace{.1in}
\caption{Tsunami  inflation:
$ h(\tau) $ for $ \tau > \tau_A $. The early time analytic
approximation gives $ h_A = 36.1 $ (also with the one mode approx.),
numerically we obtain $ h(\tau_A) = 35.5 $.
Same parameters and initial conditions as in fig. \ref{hearlyfig}.}
\label{hfig}
\end{figure}

\begin{figure}[h]
\epsfig{file=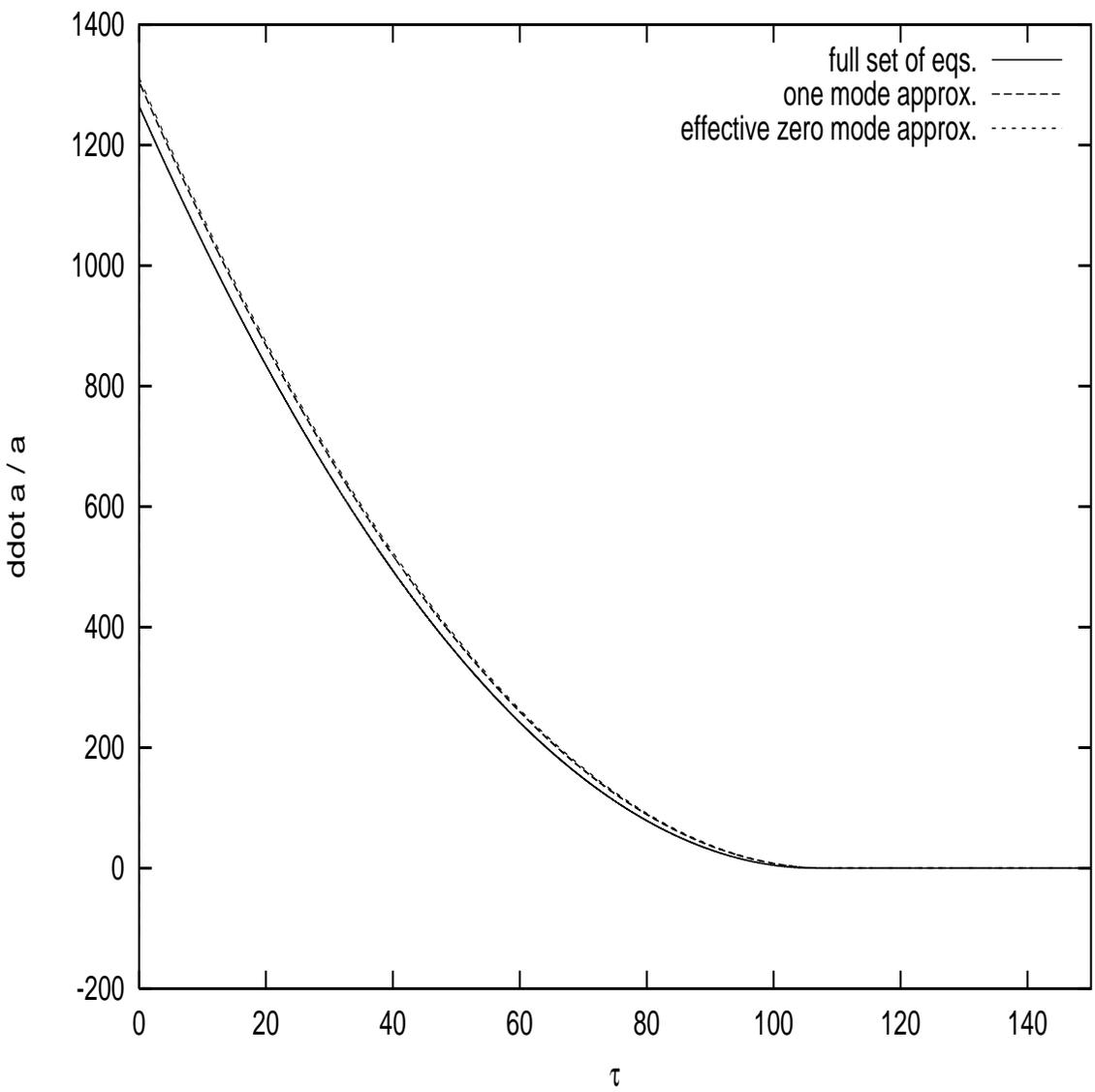,width=6in,height=6in}
\vspace{.1in}
\caption{Tsunami  inflation:
$ \ddot a(\tau) \over a(\tau) $, it shows that there is accelerated
expansion (inflation) up to times $ \tau \sim 109 $.
Same parameters and initial conditions as in fig. \ref{hearlyfig}.}
\label{ddaoverafig}
\end{figure}

\begin{figure}[h]
\epsfig{file=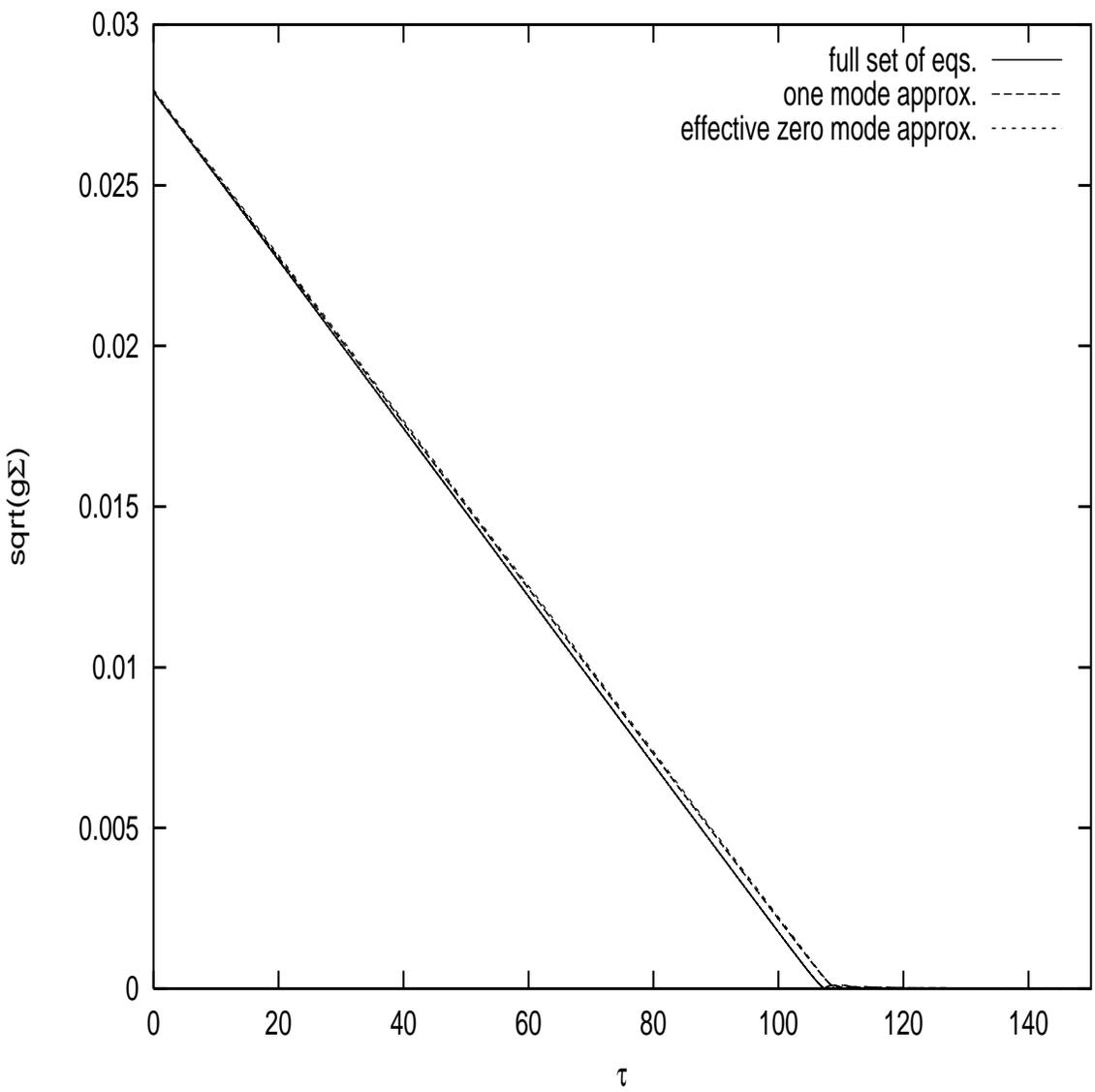,width=6in,height=6in}
\vspace{.1in}
\caption{Tsunami  inflation:
$ \sqrt{g\Sigma(\tau)} $, after $ \tau_A \sim 0.133 $, it plays
the role of an effective classical field.
Same parameters and initial conditions as in fig. \ref{hearlyfig}.}
\label{gsigmafig}
\end{figure}

\begin{figure}[h]
\epsfig{file=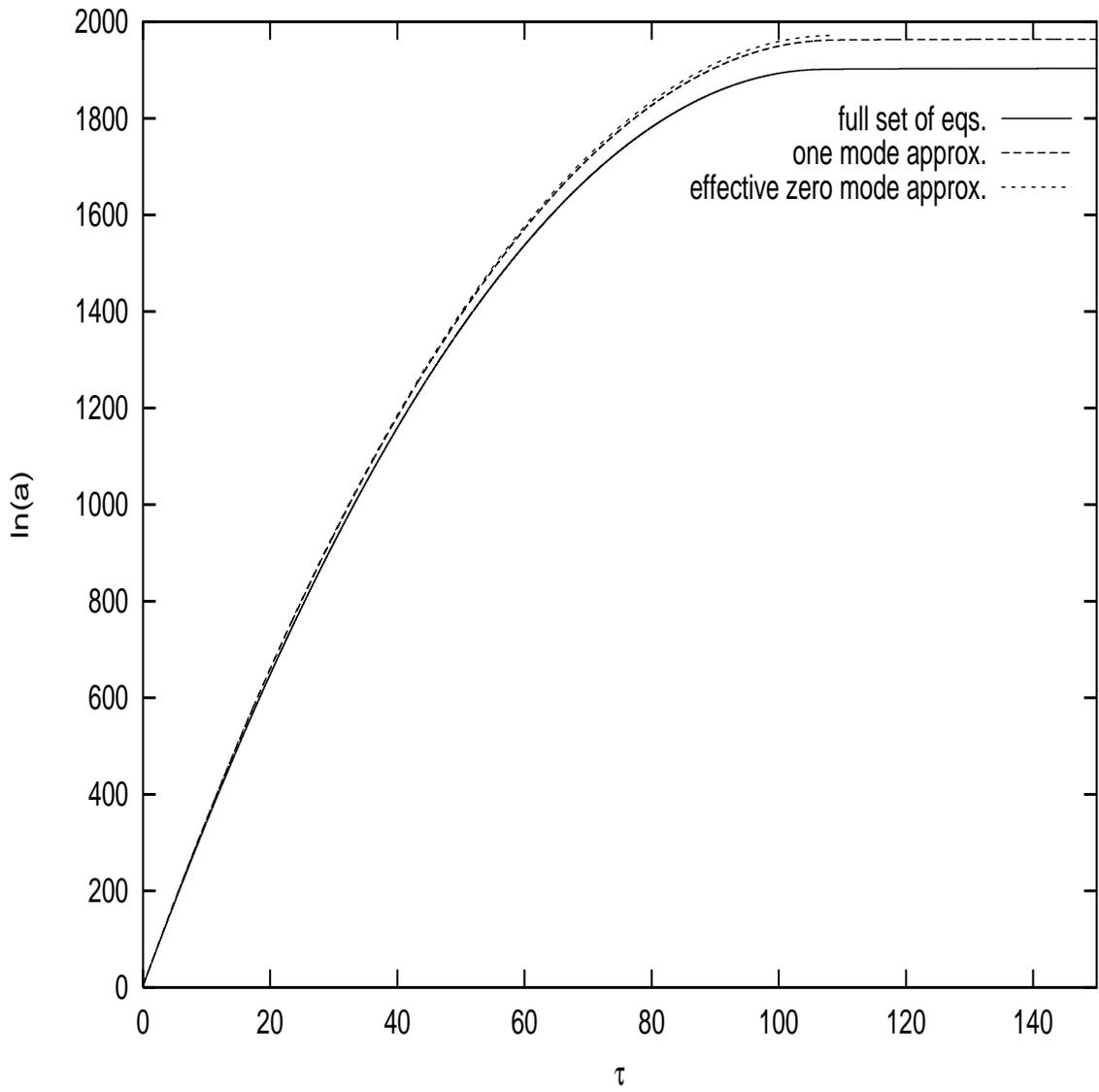,width=6in,height=6in}
\vspace{.1in}
\caption{$ \ln[a(\tau)] $ vs. $\tau$.
Same parameters and initial conditions as in fig. \ref{hearlyfig}.}
\label{lnafig}
\end{figure}

\begin{figure}[h]
\epsfig{file=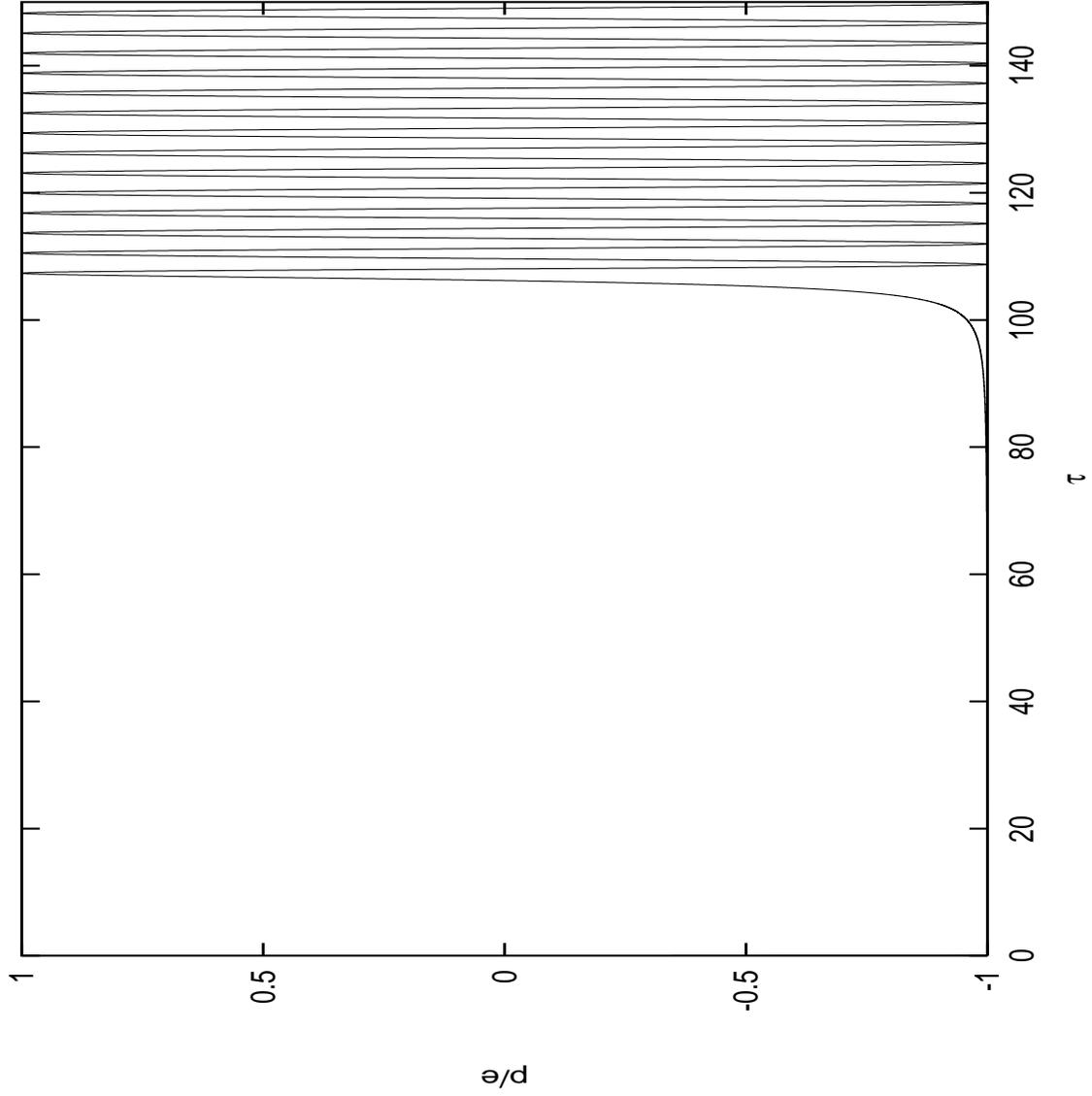,width=6in,height=6in}
\vspace{.1in}
\caption{Tsunami  inflation:
$ p(\tau)/\epsilon(\tau) $. It shows the onset of a matter dominated
epoch after the quasi-De Sitter stage.
Same parameters and initial conditions as in fig. \ref{hearlyfig}.}
\label{poverefig}
\end{figure}

Figure \ref{Nefig} displays $\epsilon_0$ vs $q_0$ along lines of
constant number of e-folds , while figures
\ref{hearlyfig}-\ref{poverefig} display the solution of the
full set of equations  (\ref{modknr})-(\ref{h2tau})
with (\ref{energydens}). An important feature that
emerges from these figures is that for the set of parameters that are typical
for inflationary scenarios and for large values of $q_0=k_0/m$ (but
well below the Planck scale) the number of e-folds obtained is more
than sufficient as shown by fig.\ref{lnafig}.

We also show that the dynamics of the full set of equations
(\ref{modknr})-(\ref{h2tau}) with
(\ref{energydens}) is correctly approximated  by the narrow tsunami
case studied in the previous subsections:
the one mode approximation [eqs. (\ref{apgsigI})-(\ref{eypq0})],
the early time analytical formulae (for $ \tau \le \tau_A $)
[eqs. (\ref{soltaua})-(\ref{ctestaua})],
and the effective classical field (for $ \tau > \tau_A $)
[eqs. (\ref{peeff})]. The agreement between the analytic treatment and
the full numerical evolution is displayed in figures
\ref{hearlyfig}-\ref{poverefig}.

The early time analytic expressions predict an inflationary
period during which the Hubble parameter falls off fairly fast, that
lasts up to
$ \tau_A \sim 0.133 $ [eq. (\ref{taua})] reaching an asymptotic value of
$ h(\tau_A) = 36.2 $ [eqs. (\ref{ctestaua}) and (\ref{valuesattaua})].
The one mode approximation gives the same prediction $ h(\tau_A) = 36.1 $,
and numerically evolving the full set of equations we find
$ h(\tau_A) =  35.5 $.
Thus, we see from this values and from figs. \ref{hearlyfig} and
\ref{hfig} that both approximations are fairly accurate  for early times.

After $ \tau_A $, the geometry reaches a quasi-De Sitter epoch. We
have shown in the
previous subsection that after the time $ \tau_A $ the evolution
equations for the
one mode approximation reduce to those of an effective classical field.
The effective zero mode approximation correctly predicts the dynamics
in this epoch as can be gleaned from figures \ref{hfig}-\ref{lnafig}.

While the stage of early tsunami inflation up to  $ \tau_A $ results in
only $ 8.5 $ efolds, the following quasi-de Sitter stage described by the
effective classical scalar field
lasts for  a total of $ 1900 $ efolds. For the values of
parameters chosen above,  $ g\Sigma_0 \ll 1 $, hence we can estimate the number
of efolds with eq. (\ref{lnane0narrow}). Using eq. (\ref{nenarrow}) we obtain
a total of $ 1970 $ efolds while
the one mode approximation yields $ 1960 $ efolds. Both results agree
with the full numerical solution of the equations (see fig. \ref{lnafig}).

Furthermore, as stated above  inflation ends when
$ h \sim \frac{1+\eta_{eff}^2}{3} \sim \frac{1}{3} $, after which  a matter
dominated epoch follows.

\subsection{Other distributions and other states:} \label{sotherdist}
The validity of the physical picture that emerges from the previous analytic
and numerical study is not restricted to
pure states or narrow distributions of the form given by
(\ref{distributions}). We have also studied more general
distributions and mixed states:

\vspace{1mm}

{\em Other distributions: }

The narrow tsunami case where a single quantum mode $q_0$
dominates the dynamics has been extremely useful to study the dynamics in the
previous section. The generalization to the case with
continuous distributions of $q$-modes can be easily obtained making
the changes:
\begin{eqnarray}\label{cambio}
|\phi_{q_0}(\tau)|^2 &\to& g \, \int q^2 \; dq \; |f_q(\tau)|^2 \; , \cr \cr
|{\dot\phi}_{q_0}(\tau)|^2 &\to& g \, \int q^2 \; dq \; |{\dot f}_q(\tau)|^2
\; , \cr \cr
q_0^2 \; |\phi_{q_0}(\tau)|^2 &\to& g \, \int q^2 \; dq \; q^2
\; |f_q(\tau)|^2 \; .
\end{eqnarray}
The two stages of inflation are always present for such continuous
modes distribution as long as the following generalized slow-roll
conditions is fulfilled
\begin{equation}\label{genslo}
g \, \int q^2 \; dq \; |{\dot f}_q(\tau)|^2 \ll
g \, \int q^2 \; dq \; |f_q(\tau)|^2
\end{equation}
that imposes on $ \delta_q $ the condition $ |\omega_q \, \delta_q +
h(0)| \ll 1 $.

The effective zero mode in the second stage of inflation is now given by
$$
{\eta}^2_{eff}(\tau) = g \, \int q^2 \; dq \; |f_q(\tau)|^2 \; .
$$
Our numerical study with general distributions reveals that the
analytical picture obtained by substituting eq.(\ref{cambio}) in
sec. IIIA correctly reproduce the dynamics.

\vspace{1mm}

{\em Other (mixed) states:} Although we have focused for simplicity on
tsunami pure initial states, we have also
investigated the possibility of mixed states. Mixed state density
matrices and their time evolution are discussed in the appendix. The
mixing can be parametrized in terms of angles $\Theta_k$ as given in
equation (\ref{defcalA}) and the number of (conformal)  quanta
are given by eq. (\ref{occumix}). The {\em only} relevant changes that
occur are  in the integrals for
$\Sigma(\tau)~;~\varepsilon(\tau)~;~p(\tau)$ in which
\be
|{f}_q(\tau)|^2 \rightarrow |{f}_q(\tau)|^2
\coth\left[\frac{\Theta_q}{2}\right]
 \quad , \quad
|\dot{f}_q(\tau)|^2 \rightarrow |\dot{f}_q(\tau)|^2
\coth\left[\frac{\Theta_q}{2}\right] \label{change}
\ee
Tsunami and slow-roll conditions on the mode functions given by
eqs. (\ref{bigomega}) and (\ref{slowrolldelta}) lead
to tsunami-wave inflation followed by chaotic inflation just as
discussed above. In the narrow
tsunami case the only change is that the effective $q_0$-mode
is rescaled by the mixing factor, i.e,
\be
|\phi_{q_0}(\tau)|^2 \rightarrow |\phi_{q_0}(\tau)|^2
\coth\left[\frac{\Theta_{q_0}}{2}\right] \quad , \quad
|{\dot\phi}_{q_0}(\tau)|^2 \rightarrow |{\dot\phi}_{q_0}(\tau)|^2
\coth\left[\frac{\Theta_{q_0}}{2}\right]
\label{mixedq0modo}
\ee
It is also illuminating to contrast the tsunami-wave mixed
states with the more familiar {\em thermal} mixed states. The latter
are obtained by the choice
\be
\Omega_q = \omega_q ~~; ~~ \delta_q = 0 ~~; ~~  \Theta_q
= \frac{\omega_q}{T} \label{thermal}
\ee
\noindent with $T$ some value of temperature. In this case it is
straightforward to see that (quantum) equipartition
results in that the contributions of the modes and their time
derivatives to the energy and  pressure are of the same order
  ($ |\dot{f}_q(\tau)|^2 \sim [ h(0)^2 +\omega_q^2] \;
|{f}_q(\tau)|^2$). Hence, for
these thermal mixed states the tsunami slow-roll condition is not fulfilled.
This is obviously {\em not} surprising, such a choice of thermal mixed
state leads to a FRW epoch which
is not inflationary. Hence the tsunami-wave initial conditions along
with the generalized slow-roll conditions lead to {\em two} successive
inflationary epochs in striking contrast to the familiar mixed thermal
states.
\section{Generalized chaotic inflation} \label{sgenchaoinf}

The previous analysis, confirmed by the numerical evolution of the
full self-consistent set of equations leads to one of the important
conclusions of this article, that tsunami-wave initial states provide
a microscopic justification of the chaotic inflationary scenario.

We have focused our discussion on initial states with vanishing
expectation value of the scalar field (order parameter) and where
the energy is concentrated in a momentum band (tsunami initial
states). This choice brings to the fore the striking contrast
between this novel {\em quantum state} and the usual classical
approach to chaotic inflation. In this section we study the
dynamics in the case in which the initial state allows for a
non-vanishing expectation value of the scalar field {\em along}
with some of the initial energy localized in excited quanta. We
refer to this case  as generalized chaotic inflation to
distinguish from the tsunami-wave state studied above. This
generalization thus includes both cases: the classical chaotic
inflation in the limit when there are no excited modes, as well as
the tsunami initial state when all of the energy is localized in a
band of excited modes and the expectation value of the field
vanishes.

The relevant equations of motion in comoving time for the mode
functions in this case are given by (\ref{equationsgene}) in the appendix.
Along with the dimensionless variables (\ref{dimvars1})  it is also
convenient to introduce a dimensionless expectation value as
\be
\eta^2(\tau) = \frac{\lambda}{2m^2} \; \phi^2(t) \label{dimlessexp}
\ee

In this generalized case with $\eta \neq 0$ the equations of motion
for the mode functions $f_q(\tau)$
(in terms of dimensionless variables) are the same as in
eq. (\ref{modknr}) after the replacement
$g\Sigma(\tau)\rightarrow g\Sigma(\tau)+\eta^2(\tau)$ and the equation
of motion for $\eta(\tau)$ is given by
\begin{eqnarray}
&& \frac{d^2 \eta({\tau})}{d\tau^2}+3h(\tau)
\frac{d\eta(\tau)}{d\tau}+\left[1+\eta^2(\tau)+g\Sigma(\tau)
\right]\eta(\tau)  =  0 \label{etacomo}  \\
&& \eta(0)  =  \eta_0 \quad ; \quad
\dot{\eta}(0)  = \dot{\eta}_0 \label{condinieta}
\end{eqnarray}
The Einstein-Friedmann equation is given by (\ref{h2tau}) but with the
energy density and pressure now given by
\begin{eqnarray}\label{enepre}
\epsilon(\tau) &  =  &  \frac{1}{2}\dot{\eta}^2+ \frac12 \,
\left(g\Sigma+\eta^2\right) +
\frac14 \, \left(g\Sigma+\eta^2\right)^2 + \nonumber \\
&+ & \frac{g}{2}\int q^2 \; dq \left\{|\dot{f_q}|^2 -
S_1(q,\tau) +\frac{q^2}{a^2}\, \left[|f_q|^2 - S_2(q,\tau)\right]
\right\} \; ,\label{enerdensgene}\\
 (p+\epsilon)(\tau)  & = &
\dot{\eta}^2  +   g \int q^2\; dq \left\{ |\dot{f_q}|^2 - S_1(q,\tau)
+\frac{q^2}{3a^2}\left[  |f_q|^2 - S_2(q,\tau) \right] \right\}
\;. \label{pmasegene}
\end{eqnarray}
\noindent where the renormalization subtractions $ S_1~;~S_2 $ are
obtained from those given by eq. (\ref{renosubs}) upon the
replacement $g\Sigma \rightarrow g\Sigma+\eta^2$.

From this expression we see that for  fixed (large) energy density
Status: RO

there are two different possibilities: if the  zero
mode squared $\eta^2(0)$ is larger than
the quantum fluctuations $ g\Sigma $ and $g \int q^4 \; dq \; |f_q|^2 $,
the dynamics is basically similar to that in the usual chaotic
inflationary scenarios. This
corresponds to most of the initial energy density to be in the zero mode and
little energy density in the band of excited states. On the other hand,
for small $\eta^2(0)$ most of the initial energy density is
in the tsunami quantum state and the initial dynamics is akin to the
$\eta=0$ case. To quantify this statement and clarify the
interplay and crossover of behaviors between the $\eta=0$ and the
generalized chaotic case, we now resort again to the narrow
tsunami case, which highlights the essential physics.

The relevant equations are: i) the equations of motion for the
effective $q_0-$mode (\ref{effzeromode}),
\begin{equation}
{\ddot \phi}_{q_0}(\tau) + 3 \, h(\tau) \, {\dot \phi}_{q_0}(\tau) + \left[
{q_0^2 \over a^2(\tau)} + 1 + \eta^2(\tau)+|\phi_{q_0}(\tau)|^2
\right]\phi_{q_0}(\tau) = 0 \; .\label{modoq0gen}
\end{equation}
\noindent and  for the  zero mode $\eta$
\be
{\ddot \eta}(\tau)+3h(\tau)
{\dot \eta}(\tau)+\left[1+\eta^2(\tau)+|\phi_{q_0}(\tau)|^2
\right]\eta(\tau)  =  0 \label{etacomogen}
\ee
\noindent and the Hubble parameter given by (\ref{friedq0}) with the
energy density given by
\be
\epsilon(\tau) = \frac12 \, \left(|{\dot \phi}_{q_0}(\tau)|^2 +{\dot
\eta}^2(\tau)\right)
+ \frac12 \, \left(|\phi_{q_0}(\tau)|^2+\eta^2(\tau)\right) + \frac14
\; \left(|\phi_{q_0}(\tau)|^2 +\eta^2(\tau)\right)^2
+ \frac{q_0^2}{2 \, a^2(\tau)} \; |\phi_{q_0}(\tau)|^2 .\label{eypq0gen}
\ee
The acceleration of the scale factor is now given by
\begin{equation}
\frac{\ddot a(\tau)}{a(\tau)} = - L^2 \left[|{\dot
\phi}_{q_0}(\tau)|^2+{\dot \eta}^2(\tau) - \frac{1}{2}\left(
|\phi_{q_0}(\tau)|^2+\eta^2(\tau)\right)-
\frac{1}{4}\left(|\phi_{q_0}(\tau)|^2+\eta^2(\tau)\right)^2  \right]
\label{accgen}
\end{equation}

\noindent where again we have neglected the renormalization contributions and terms of ${\cal O}(g)$
 consistently in the weak coupling limit $g\ll 1$. From
eq. (\ref{accgen}) the generalized condition for an inflationary
epoch (within the  narrow tsunami case) becomes
\be
|{\dot \phi}_{q_0}(\tau)|^2+{\dot \eta}^2(\tau) < \frac{1}{2}\left(
|\phi_{q_0}(\tau)|^2+\eta^2(\tau)\right)+
\frac{1}{4}\left(|\phi_{q_0}(\tau)|^2+\eta^2(\tau)\right)^2
\ee
\noindent which is fulfilled if the following {\em generalized slow
roll condition} holds
\be
|{\dot \phi}_{q_0}(\tau)|^2+{\dot \eta}^2(\tau) \ll
|\phi_{q_0}(\tau)|^2+\eta^2(\tau) \label{slowrollgen}
\ee
Under these conditions and from eqs.(\ref{modoq0gen})-(\ref{etacomogen})
and the dynamics of the scale factor driven by the
energy density eq.(\ref{eypq0gen})  we can now distinguish the following
different inflationary scenarios.

\begin{itemize}

\item{{\em Tsunami dominated:} When $ (q^2_0 +1) \;
|\phi_{q_0}(0)|^2 \gtrsim \eta^2(0)$ the excited states
in the tsunami-wave carry most of the initial energy density. In this
case the results of the previous section apply and the
scale factor takes the form as in eq.(\ref{friango})  with $D$ given
by eq.(\ref{ctestaua}), and $E$
given by eq.(\ref{ctestaua}) but with $g\Sigma_0 \rightarrow
g\Sigma_0+\eta^2(0)$. There are {\em two} consecutive inflationary
stages as in the previous section. The first described by
eq.(\ref{soltaua}), lasts  up to the time scale $\tau_A$ defined by
$$
{\dot \eta}^2(\tau_A) + |{\dot \phi}_{q_0}(\tau_A)|^2 \sim
\frac{q_0^2}{3 \, a^2(\tau_A)} \; |\phi_{q_0}(\tau_A)|^2
$$
at which the red-shift of the momentum $q_0$ is such that $q_0/a(\tau_A)\ll
1$. The  secondary stage is a  usual classical chaotic inflationary
epoch determined by the dynamics of an effective zero mode given by
\be
\eta^2_{eff}(\tau) = \eta^2(\tau)+ |\phi_{q_0}(\tau)|^2 \label{etaeffgen}
\ee
\noindent because for $\tau > \tau_A$, $q^2_0/a^2(\tau)\ll 1 $ and the
effective equation of motion for $\phi_{q_0}(\tau)$ is the same as
that for $\eta(\tau)$. }

\item{{\em Zero mode dominated:} When $\eta^2(0) \gg (q^2_0 +1) \;
|\phi_{q_0}(0)|^2 $ the energy density stored in the
zero mode is much larger than that contributed by the excited states in the
tsunami-wave. In this case the energy density  eq.(\ref{eypq0gen}) is
completely dominated by the zero mode. The ensuing dynamics is the
familiar  classical chaotic scenario driven by a classical zero mode, {\em
without} an early stage in which the scale factor is given by
eq.(\ref{soltaua}) which is the hallmark of the tsunami-wave dynamics.}

\end{itemize}

This analysis in the narrow tsunami case does highlight the
important aspects of the dynamics in a clear manner, allowing
a clean separation of the two cases described above. We have carried a
full numerical integration of the equations of motion that
reproduce the results described above. The criterion for the crossover
between tsunami-wave and classical chaotic inflation is determined by
the relative contributions to the energy density from  the quantum
fluctuations in the tsunami wave state as compared to the energy
density of the zero mode .

The previous results [eqs.(\ref{dimlessexp})-(\ref{etaeffgen})] can be easily
generalized for generic continuous distributions of modes and for
mixed states. One has just to make the changes indicated in
eq.(\ref{cambio}) for generic distributions and in eq.(\ref{change})
for mixed states.

The generalized slow-roll conditions takes then the form:
$$
{\dot \eta}^2(\tau) + g \, \int q^2 \; dq \; |{\dot f}_q(\tau)|^2 \,
 \coth\left[\frac{\Theta_q}{2}\right]\ll
 {\eta}^2(\tau) + g \, \int q^2 \; dq \; |f_q(\tau)|^2 \,
 \coth\left[\frac{\Theta_q}{2}\right]
$$
during the first stage of inflation.

The dynamics is tsunami dominated provided,
$$
g \, \int q^2 \; dq \; (1 + q^2) \, |f_q(0)|^2 \,
\coth\left[\frac{\Theta_q}{2}\right]  \gtrsim \eta^2(0)
$$
The effective zero mode in the second stage of inflation is now given by
$$
{\eta}^2_{eff}(\tau) = {\eta}^2(\tau) + g \, \int q^2 \; dq \;
|f_q(\tau)|^2 \, \coth\left[\frac{\Theta_q}{2}\right]\; .
$$
These results have been verified by numerical integration of the full
set of evolution equations (\ref{pmase})-(\ref{h2tau}).

\section{Conclusions}

In this article we have studied inflation in typical scalar field
theories as a consequence of the time evolution of a novel quantum
state. This quantum state is characterized by a {\em vanishing}
expectation value of the scalar field, i.e,
a vanishing zero mode, but  a non-perturbatively large number of
quanta in a momentum band, thus its name--tsunami-wave state.

This state leads to a non-perturbatively large energy density which
is localized in the band of excited quantum modes.
We find that the self-consistent equations for the evolution of this
quantum state and the scale factor lead to inflation under
conditions that are the quantum analog of slow-roll.

The self-consistent evolution was studied analytically and numerically
in a wide range of parameters for the shape and position of the
distribution of excited quanta. The numerical results confirm all the
features obtained from the analytic treatment.

Under the conditions that guarantee inflation,  there are two
consecutive but distinct
inflationary epochs. The first stage features a rapid fall-off of the
Hubble parameter and is characterized by
the quantum aspects of the state. During this first stage the large
number of quanta in the excited band are redshifted and build up
an {\em effective homogeneous classical condensate}. The amplitude of
this condensate is non-perturbatively large, of ${\cal O}(1/\lambda)$,
as a consequence of the non-perturbatively large number of quanta in
the band of excited modes.

The second stage is similar to the classical chaotic scenario and it is
driven by the dynamics
of this effective classical condensate, with  vanishing  expectation
value of the scalar field. Under the tsunami slow-roll
conditions  on the quantum state, the total number of e-folds is more
than enough to satisfy the constraints of
inflationary cosmology. The band of excited wave-vectors if not
initially outside the causal horizon, becomes superhorizon
during the first inflationary stage, therefore these excited states do
not modify the power spectrum of scalar
density perturbations  on wavelengths that are of cosmological
relevance today.

Therefore, these tsunami-wave quantum states provide a quantum field
theoretical justification of chaotic (or in
general large field) inflationary models and yield to a
microscopic understanding of the emergence of classical
homogeneous field configurations of large amplitude as an effective
collective mode built from the large number
of quanta in the excited band.

In addition, we recall that it is necessary to choose an initial state
that breaks the $ \Phi \to - \Phi $ symmetry in classical chaotic
scenarios \cite{revius,coles}. This is {\em not} the case here. We
have inflation with {\em zero} expectation value of the scalar field.

For completeness we have also studied more general states and
established the important difference between tsunami (pure or mixed)
quantum states leading to inflation, and thermal mixed states which do
not lead to inflation.

Acknowledgements:
D.B. thanks NSF for support through grants PHY-9605186, PHY-9988720
and  NSF-INT-9815064.
H. J. d. V. thanks the CNRS-NSF collaboration for support. F. J. C. thanks the
Ministerio de Educaci\'on y Cultura (Spain) for financial support through
the F. P. U. Programme.
\appendix

\section{Equations of motion in the large $ N $ limit and initial states.}

 In this appendix we obtain the equations of motion in conformal time
for the generalized case in which
the initial state is determined by a density matrix. The evolution of
the functional density matrix is given by the Liouville
equation in conformal time
\begin{eqnarray} \label{Liouveq}
i\frac{\partial\rho}{\partial{\cal T}} &=& [H,\,\rho]\quad\Longrightarrow\cr\cr
i\frac{\partial\;}{\partial {\cal T}} \,
\rho[\vec\Psi,\,\vec{\tilde{\Psi}};\,{\cal T}] &=&
\left( H \left[ \frac{\partial\;}{\partial\vec\Psi};\,\vec\Psi \right]
- H \left[ \frac{\partial\;}{\partial\vec{\tilde{\Psi}}};
\,\vec{\tilde{\Psi}} \right] \right)\;
\rho[\vec\Psi,\,\vec{\tilde{\Psi}};\,{\cal T}]
\end{eqnarray}

\noindent where the Hamiltonian $H$ is given by eq.(\ref{hamk}) to
leading order in the large $N$ limit. Consistently with
the fact that in the large $N$ limit the Hamiltonian describes a
collection of harmonic oscillators, we propose
a  Gaussian density matrix
\begin{equation} \label{defrho}
\rho[\vec\pi, \vec{\tilde{\pi}}, {\cal T}] = {\cal N}({\cal T})
  \prod_{k}\exp\left\{-\frac{A_{k}({\cal T})}{2}\;
    \vec\pi_{{k}}\cdot\vec\pi_{-{k}}
  -\frac{A_{k}^*({\cal T})}{2}\;
    \vec{\tilde{\pi}}_{{k}}\cdot\vec{\tilde{\pi}}_{-{k}}
  -B_{k}({\cal T})\;\vec\pi_{{k}}\cdot
  \vec{\tilde{\pi}}_{-{k}} \right\}
\end{equation}
The hermiticity condition $ \rho^\dag = \rho $ for the density matrix
impose that $ B_{k} $ must be real.
In addition, since $ \vec\pi({\bf x}, {\cal T}) $ is a real field, its Fourier components must
obey the hermiticity condition
$ \vec\pi_{-{k}}({\cal T})= \vec\pi^*_{{k}}({\cal T}) $;
thus, we can assume $ A_{-{k}}({\cal T}) = A_{k}({\cal T}) $
without loss of generality.

The evolution equations for $ A_{k}({\cal T}) $, $ {\cal N}({\cal T}) $
and $ B_k({\cal T}) $ are obtained from the Liouville eq. (\ref{Liouveq})
where the hamiltonian is given by eq. (\ref{hamk}). We find
\begin{eqnarray}
i A^{'}_{k} &=& A^2_{k} -  B^2_{k}
  - {a^2({\cal T})}\,  \omega^2_k({\cal T}) \qquad , \qquad
i B^{'}_{k} = B_{k}\,(A_{k} - A^*_{k})
 \cr\cr
{\cal N}_{\rho}({\cal T}) & = & {\cal N}_{\rho}(0) \; e^{-{i N \over
2}\int_0^{\cal T} d{\tilde{ \cal T}} \sum_{k}
\left[A_{k}({\tilde{ \cal T}}) - A^*_{k}({\tilde{ \cal T}})\right] }
\; , \label{norma}
\end{eqnarray}
where the prime denotes derivative with respect to conformal time $ {\cal T} $.

The normalization factor for mixed states $ {\cal N}_{\rho}({\cal T})
$ is related with the normalization factor of pure states $ {\cal
N}_{\Upsilon}({\cal T}) $ by
$$
{\cal N}_{\rho}({\cal T}) = {\cal N}_{\Upsilon}({\cal T}) \; {\cal
N}_{\Upsilon}({\cal T})^*
$$
where
$$
{\cal N}_{\Upsilon}({\cal T})= {\cal N}_{\Upsilon}(0) \;
\exp\left\{-i\int^{\cal T}_0{d{\cal T}'
\left[  N V h_{cl}({\cal T}') - \frac{\lambda}{8 \, N}
\left( \sum_{k} \langle \vec\pi_{{k}}\cdot\vec\pi_{-{k}}
\rangle({\cal T}')\right)^2 + {N \over 2} \sum_{k}A_{k}({\cal T}') \right]}
\right\}
$$

Writing $ A_{k} $ in terms of its real and imaginary parts
$ A_{k} = A_{R,{k}} + i A_{I,{k}} $, we find that $ B_{k}/A_{R,{k}} $
is a conserved quantity. Thus, we can introduce  without loss of
generality the  variables $  {\cal A}_{R,{k}}({\cal T}), \;  {\cal
A}_{I,{\bf k}}({\cal T}) $ and $ \Theta_{k} $ defined by
\begin{eqnarray} \label{defcalA}
A_{R,{k}}({\cal T}) &\equiv& {\cal A}_{R,{k}}({\cal T})
\,\coth\Theta_{k} \quad , \quad A_{I,{k}}({\cal T}) \equiv {\cal
A}_{I,{k}}({\cal T}) \cr\cr
B_{k}({\cal T})&\equiv& -\;\frac{{\cal A}_{R,{k}}({\cal T})}{\sinh\Theta_{k}}
\end{eqnarray}
where $ \Theta_{k} $ is a time independent  real function.

Introducing the complex variable
\begin{equation}
{\cal A}_{k} = {\cal A}_{R,{k}} + i {\cal A}_{I,{k}}
\end{equation}
we see that it obeys the following Ricatti equation
\begin{equation} \label{riccalA}
i {\cal A}^{'}_{k} = {\cal A}^2_{k}
 - a^2({\cal T})\; \omega^2_k({\cal T})
\end{equation}
This equation can be linearized defining
\begin{equation} \label{defphik}
{\cal A}_{k}({\cal T}) \equiv - i\,
  \frac{\varphi^{'*}_{k}({\cal T})}{\varphi^*_{k}({\cal T})} \; .
\end{equation}
Then eq. (\ref{riccalA}) implies that the mode functions $ \varphi_{k} $
obey
\begin{eqnarray} \label{eqevolmodconf}
&&\varphi^{''}_{k} +  \omega^2_ k({\cal T})\; \varphi_{k}
= 0 \; , \cr\cr
&& \omega^2_{ k}({\cal T}) = k^2 + a^2({\cal T})\, \left[ {\cal M}^2({\cal
T}) - \frac{{\cal R}({\cal T})}{6} \right]   \label{omegadef}\; ,
\end{eqnarray}
where $ {\cal R}({\cal T}) $ is the Ricci scalar.

The relation (\ref{defphik}) defines the mode functions $
\varphi_{k}({\cal T}) $ up to an arbitrary multiplicative constant
that we choose such that the wronskian takes the value,
\begin{equation} \label{wrons}
\varphi_{k}\,\varphi^{'*}_{k} - \varphi^{'}_{k}\,\varphi^*_{k} = 2 i \; .
\end{equation}
For this choice of the Wronskian the definition (\ref{defphik}) becomes
\begin{equation}
{\cal A}_{k} = \frac{1}{|\varphi_{k}|^2}
 -\frac{i}{2}\, \frac{d\;}{d{\cal T}}\,\ln|\varphi_{k}|^2  \; .
\end{equation}
The mass term in eq.(\ref{omegadef}) given by eq. (\ref{M2}) requires the
self-consistent expectation value
\begin{eqnarray}
\frac{\langle\vec\pi^2\rangle_\rho}{N} &=& \int{\frac{d^3k}{(2\pi)^3}\,
  \langle\vec\pi_{{k}}\cdot\vec\pi_{-{k}}\rangle_\rho}
\cr\cr
\langle\vec\pi_{{k}}\cdot\vec\pi_{-{k}}\rangle_\rho
  &=& \mbox{Tr}\rho \,
  \vec\pi_{{k}}\cdot\vec\pi_{-{k}}=\frac{1}{2\,[A_{R,{k}}+B_{k}]}
  = \frac{1}{2\,{\cal A}_{R,{k}}}\,
  \coth\left(\frac{\Theta_{k}}{2}\right)
  = \frac12 \; |\varphi_{k}|^2\,
  \coth\left(\frac{\Theta_{k}}{2}\right)
\end{eqnarray}
Thus, the evolution equations in terms of the mode functions are given
by eq. (\ref{eqevolmodconf}) with
\begin{equation} \label{explmass}
{\cal M}^2({\cal T}) = m^2 + \xi\,{\cal R}
  + \frac{\lambda}{2}\,\frac{\psi^2}{a^2}
  + \frac{\lambda}{4}\,\int{\frac{d^3k}{(2\pi)^3}\,
  \frac{|\varphi_{k}|^2}{a^2}\, \coth\left(\frac{\Theta_{k}}{2}\right)}
\end{equation}
The evolution equation of the mode  functions $ \varphi_{k} $ is the
same as  the Heisenberg equations of motion
for the fields, hence we can write the Heisenberg field operators as
\begin{equation}
\vec\pi({\bf x},\,{\cal T})
  = \int{\frac{d^3k}{\sqrt2(2\pi)^3} \, \left[
  \vec{a}_{k} \; \varphi_{k}({\cal T}) \; e^{i\,{k}\cdot{\bf x}}
  + \vec{a}^\dag_{k}\,\varphi^*_{k}({\cal T})\,e^{-i\,{k}\cdot{\bf x}}
  \right]}
\end{equation}
Thus, the definition (\ref{defphik}) gives the relation between
Schr\"odinger and Heisenberg pictures, since the functional density matrix
(\ref{defrho}) is in the Schr\"odinger picture.

The expectation value $\psi({\cal T})$ [see eq. (\ref{fieldsplit})] in
conformal time obeys the following equation of motion\cite{noscos}
\bea
&&\psi^{''}({\cal T}) + a^2({\cal T})\, \left[ {\cal M}^2({\cal T})
  - \frac{{\cal R}({\cal T})}{6} \right]\;\psi({\cal T}) = 0 \label{evoleqexpconf}\\
&&\psi(0)=\psi_0 ~~;~~ \psi'(0)=\psi'_0 \label{iniconexpval}
\eea

Hence, the evolution equations  are given by
(\ref{eqevolmodconf}), (\ref{explmass}) and (\ref{evoleqexpconf}) with (\ref{iniconexpval}).

The initial density matrix in the Schr\"odinguer
picture is  determined by  specifying the initial values of $ {\cal A}_{R,{ k}} $,
$ {\cal A}_{I,{ k}} $ and $ \Theta_{ k} $. We will take $ a(0) = 1 $
and parameterize the initial value of $ {\cal A}_{{ k}} $ as follows,
\begin{equation} \label{schinicond}
{\cal A}_{R,{ k}}(0) = \Omega_{ k} \quad , \quad
{\cal A}_{I,{ k}}(0) = \omega_k(0)\; \delta_{ k}
\end{equation}
The corresponding initial conditions for the mode functions
 are obtained from eq.(\ref{schinicond}) using eq.(\ref{defphik}) and
 the Wronskian constraint  eq.(\ref{wrons}). These are given by
\begin{equation} \label{inimodeconf}
\varphi_{ k}(0) = \frac{1}{\sqrt{\Omega_{k}}} \;; \quad\quad
\varphi^{'}_{ k}(0) = - [\omega_k(0)\,\delta_{ k}
+ i \Omega_{ k}] \; \varphi_{ k}(0)
\end{equation}
Defining the number of particles in terms of the adiabatic eigenstates
of the Hamiltonian (\ref{hamk}) as in
eq. (\ref{ocunumb}), it is straightforward to find that the  initial
occupation numbers are given by
\begin{equation}
n_{k}(0) = \langle \hat{n}_{ k}(0) \rangle_{\rho(0)} =
\frac{\Omega_{ k}^2 + \omega^2_{k}(0)
+ \omega^2_{ k}(0)\,\delta^2_{k}}{4\,\omega_k\,\Omega_{ k}}\,
\mbox{coth}\left(\frac{\Theta_{ k}}{2}\right)-\frac12 \label{occumix}
\end{equation}
For any mixing parameter $\Theta_k \neq 0$ the density matrix
represents a {\em mixed state} since $B_k \neq 0$, a {\em pure}
initial state is obtained by  taking $\Theta_{k} = \infty$, in which
case $ B_k \rightarrow 0 $ and the density matrix becomes a
product of a wave functional times its complex conjugate.

It is convenient to pass to comoving time, this is achieved by the
rescaling of the fields
\be
\psi({\cal T}(t)) = \phi(t) \; a(t)\quad , \quad
\varphi_{k}({\cal T}(t))= f_k(t) \; a(t) \label{zeromodecomo}
\ee
\noindent in terms of which  the equations of motion are
\begin{eqnarray}
&&\ddot \phi(t) + 3\,H(t)\,\dot\phi(t) + {\cal M}^2(t) \phi(t) = 0 \cr\cr
&&\ddot f_{k}(t) + 3\,H(t)\,\dot f_{k}(t) + \left[ \frac{k^2}{a^2(t)}
+ {\cal M}^2 (t)    \right] f_{k}(t) = 0 \cr\cr
&&{\cal M}^2(t) = m^2 + \xi\,{\cal R}(t) + \frac{\lambda}{2}\,\phi^2(t)
  + \frac{\lambda}{4}\,\int{\frac{d^3k}{(2\pi)^3}\,
  |f_{k}(t)|^2\, \coth\left(\frac{\Theta_{k}}{2}\right)} \label{equationsgene}
\end{eqnarray}
where the dots denote derivative with respect to the comoving time
$t$. The initial conditions for the order parameter are its initial value $
\phi(0) $, and its initial derivative $ \dot\phi(0) $. For $ a(0) = 1 $,
the initial conditions for the fluctuations are given by $ \Theta_{k} $ and
\begin{equation} \label{inicondmodcom}
f_{k}(0) =  \frac{1}{\sqrt{\Omega_{k}}} \;; \quad \quad
\dot f_{k}(0) =  - [\omega_k(0)\,\delta_{k} + H(0)
+ i \Omega_{k}] \; f_{k}(0)
\end{equation}
[Those are the transformed of the initial conditions in conformal time
eq. (\ref{inimodeconf}).]


\end{document}